\documentclass[10pt]{iopart}
\usepackage{graphicx,epsfig}
\usepackage{bm}
\usepackage{iopams}
\newcommand{\ba}{\begin{eqnarray}}
\newcommand{\ea}{\end{eqnarray}}
\newcommand{\bse}{\numparts}
\newcommand{\ese}{\endnumparts}
\newcommand{\ACal}{{\cal{A}}}

\newcommand{\bbq}{\begin{quote}}
\newcommand{\eeq}{\end{quote}}
\newcommand{\tbb}{t_{\textrm{\tiny{bb}}}}

\newcommand{\tbbo}{t_{\textrm{\tiny{bb(0)}}}}
\newcommand{\tcoll}{t_{\textrm{\tiny{coll}}}}
\newcommand{\Tqcoll}{\tau_{q\tiny{\textrm{coll}}}}
\newcommand{\tmax}{t_{\textrm{\tiny{max}}}}
\newcommand{\Tqmax}{\tau_{q\tiny{\textrm{max}}}}
\newcommand{\amax}{a_{\textrm{\tiny{max}}}}

\newcommand{\RR}{{\cal{R}}^{(3)}}
\newcommand{\RRR}{{\cal{R}}}

\newcommand{\JJ}{{\cal{J}}}

\newcommand{\HH}{{\cal{H}}}
\newcommand{\KK}{{\cal{K}}}
\newcommand{\PP}{{\cal{P}}}

\newcommand{\UU}{{\cal{U}}}

\newcommand{\Da}{\delta^{(A)}}

\newcommand{\Dh}{\delta^{(\HH)}}

\newcommand{\Dbe}{\delta^{(\beta)}}
\newcommand{\Dal}{\delta^{(\gamma)}}
\newcommand{\DT}{\delta^{(\tau)}}

\newcommand{\Dg}{\Delta^{\tiny{\textrm{(g)}}}}
\newcommand{\Dd}{\Delta^{\tiny{\textrm{(d)}}}}

\newcommand{\Drho}{\delta^{(\rho)}}

\newcommand{\DKK}{\delta^{(\KK)}}
\newcommand{\DOm}{\delta^{(\Omega)}}
\newcommand{\dd}{{\rm{d}}}

\newcommand{\Dig}{\Delta_0^{\tiny{\textrm{(g)}}}}
\newcommand{\Did}{\Delta_0^{\tiny{\textrm{(d)}}}}

\newcommand{\jg}{J_{\tiny{\textrm{(g)}}}}
\newcommand{\jd}{J_{\tiny{\textrm{(d)}}}}
\newcommand{\Jg}{\JJ_{\tiny{\textrm{(g)}}}}
\newcommand{\Jd}{\JJ_{\tiny{\textrm{(d)}}}}

\usepackage{pdfsync}
\begin{document}

\title[Invariant characterization of the growing and decaying density modes in LTB models.]{Invariant characterization of the growing and decaying density modes in LTB dust models.} 
\author{ Roberto A. Sussman}
\address{Instituto de Ciencias Nucleares, Universidad Nacional Aut\'onoma de M\'exico (ICN-UNAM),
A. P. 70--543, 04510 M\'exico D. F., M\'exico.}
\eads{$^\ddagger$\mailto{sussman@nucleares.unam.mx}}
\date{\today}
\begin{abstract} We obtain covariant expressions that generalize the growing and decaying density modes of linear perturbation theory of dust sources by means of the exact density perturbation from the formalism of quasi--local scalars associated to weighted proper volume averages in LTB dust models. The relation between these density modes and theoretical properties of generic LTB models is thoroughly studied by looking at the evolution of the models through a dynamical system whose phase space is parametrized by variables directly related to the modes themselves. The conditions for absence of shell crossings and sign conditions on the modes become interrelated fluid flow preserved constraints that define sub--cases of LTB models as phase space invariant subspaces. In the general case (both density modes being nonzero) the evolution of phase space trajectories exhibits the expected dominance of the decaying/growing in the early/late evolution times defined by past/future attractors characterized by asymptotic density inhomogeneity. In particular, the growing mode is also dominant for collapsing layers that terminate in a future attractor associated with a ``Big Crunch'' singularity, which is qualitatively different from the past attractor marking the ``Big Bang''. Suppression of the decaying mode modifies the early time evolution, with phase space trajectories emerging from an Einstein--de Sitter past attractor associated with homogeneous conditions. Suppression of the growing mode modifies the late time evolution as phase space trajectories terminate in future attractors associated with homogeneous states. General results are obtained relating the signs of the density modes and the type of asymptotic density profile (clump or void). A critical review is given of previous attempts in the literature to define these density modes for LTB models.                                                 
\end{abstract}

\maketitle
\section{Introduction.}

The spherically symmetric Lema\^{\i}tre--Tolman--Bondi (LTB) dust models \cite{LTB} are a valuable tool to examine  non--linear non--perturbative relativistic effects of cosmological and astrophysical self--gravitating systems by means of mathematically tractable methods (see the comprehensive reviews in \cite{kras1,kras2,BKHC2009,celerier,focus}). They have been used as toy models in a wide variety of contexts: structure formation and late time cosmological inhomogeneities \cite{KH,ltbstuff}, ``void models'' fitting cosmological observations without resorting to dark energy (see reviews in \cite{BKHC2009,marranot}), testing averaging formalisms \cite{LTBave1,LTBave2,sussBR,sussIU,suss2011}, cosmic censorship \cite{lemos,joshi} and even in quantum gravity \cite{quantum}.

The standard original set of metric variables is still used in most applications, as can be seen in the various book reviews \cite{kras1,kras2,BKHC2009}, though void models often use different forms of ``FLRW lookalike'' variables that generalize FLRW observable parameters. Other alternative variables are the ``quasi--local'' (or ``q--scalars''), associated with the weighted proper volume average of covariant scalars on comoving domains \cite{part1,part2}, and successfully used in previous literature to undertake various theoretical issues: a dynamical systems approach to the models \cite{sussDS1,sussDS2}, the asymptotic behavior of covariant scalars in the radial direction \cite{RadAs}, the evolution of radial profiles and void formation \cite{RadProfs}, back--reaction and ``effective'' acceleration in the context of Buchert's formalism \cite{sussBR,sussIU,suss2011} and even to study dark energy sources compatible with the LTB metric \cite{sussQL,suss2009}. As shown in \cite{part1,part2}, the ``FLRW lookalike'' variables in void models are q--scalars, and the latter together with their associated fluctuations and perturbations are coordinate independent objects related to curvature and kinematic scalars, providing as well a complete representation of the dynamics of the models as ``exact perturbations'' on an FLRW abstract background defined by the q--scalars themselves (which satisfy FLRW time evolution laws). 

An essential feature in linear perturbation theory of dust sources is the identification of decaying and growing density modes that are, respectively, dynamically dominant in the early and late stages of the evolution. An exact non--linear generalization of these perturbation modes were first obtained for LTB models by Silk  \cite{silk} and re--derived by Krasinski and Plebanski \cite{kras2}. However, these authors merely identified special initial conditions that define the ``amplitudes'' of the modes (and allow to ``switch'' them on or off), without carrying on any further analysis. More recently, Wainwright and Andrews \cite{wainwright} obtained expressions for the density modes by attempting to express a metric function as the sum of these modes along the lines of the ``Goode-Wainwright'' variables of Szekeres models. However, their key results are misleading because these authors did not consider fully general LTB models. A critical review of all this literature is given in Appendix A. 

In the present article we extend and enhance all previous work described and summarized in the previous two paragraphs  by taking advantage of the fact that the q--scalar perturbations define a self--consistent covariant perturbation formalism \cite{part2} that is analogous (and fully equivalent, as far as LTB models are concerned) to the 1+3 perturbation formalism of Ellis, Bruni, Dunsby and van Elst \cite{ellisbruni89, BDE, LRS, 1plus3, zibin, dunsbyetal}. Therefore, instead of the metric ansatz used in \cite{wainwright}, the density growing and decaying modes are obtained from the q--scalar exact density perturbation, which has a clear covariant meaning in terms of the invariant ratio of Weyl to Ricci scalar scalar curvatures \cite{part1}. The resulting expressions for the modes are exact, fully analytic and coordinate independent,  and their ``amplitudes'' are quantities conserved by the fluid flow. These expressions are also far less complicated and easier to handle than the expressions based on the fractional comoving density gradient that were used in previous work \cite{kras2,silk,wainwright}. As a consequence, the density modes are useful to achieve a deeper understanding of various important theoretical features of the models, such as analytic solutions, simultaneity of the big bang singularity, regularity conditions for absence of shell crossings and radial profiles of the density and other covariant scalars. The effects of these modes in the dynamics of the models is best examined through a suitable dynamical system approach that unravels the connection between early/late time asymptotic inhomogeneous states and the dominance of the decaying/growing mode. This dynamical system also allows us to examine the dynamical effects of the suppression of either mode, showing that suppression of the decaying (or growing) mode leads to an invariant subspace characterized by asymptotic early (or late) time homogeneous states.   

The section by section content of the article is given as follows. We provide in sections 2 and 3 the basic necessary background material: the description of LTB models in terms of q--scalars and their perturbations defined in an initial value parametrization of the metric and the analytic solutions expressed as constraints linking the proper time length along comoving worldlines (a q--scalar in itself) and any two basic q--scalars. By considering the exact analytic form of the density perturbation, we obtain in section 4 exact coordinate independent expressions for the density growing and decaying modes, which are shown in section 5 to reduce to the familiar expressions of linear perturbations theory in the linear limit. We introduce in section 6 a dynamical system such that the evolution of the models follows as trajectories in a 3--dimensional phase space parametrized by  bounded variables directly related to the modes themselves. We analyze  in sections 7 and 8 the phase space evolution of hyperbolic and elliptic models in the general case when both modes are nonzero, while the cases when the decaying or growing modes are suppressed are examined in sections 9 and 10. In sections 11 and 12 we examine the connection between the inhomogeneity of the models (as deviation from FLRW conditions) and the phase space evolution of the perturbations and invariant curvature and kinematic scalars associated with them. Appendix A provides a  critical review of previous literature on the density modes \cite{kras2,silk,wainwright} and in Appendix B we provide the forms of the metric and basic dynamical variables in terms of the standard original variables of the models.         

\section{LTB models, q--scalars and their perturbations.}

We shall describe LTB dust models in the following useful FLRW--like metric parametrization
\ba \dd s^2 =\dd t^2+ a^2\left[\frac{\Gamma^2}{1-\KK_{q0}r^2}\dd r^2+r^2\left(\dd\vartheta^2+\sin^2\theta\dd\varphi^2\right)\right],\label{ltb2}\\
a=a(t,r),\qquad \Gamma=1+\frac{ra'}{a},\qquad a' =\frac{\partial a}{\partial r},\label{aGdef}\ea 
where $a$ satisfies the Friedman equation (\ref{HHq}), $\KK_{q0}=\KK_q(t_0,r)$ is defined further ahead (see equation (\ref{KKq})) and the subindex ${}_0$ will denote henceforth evaluation at an arbitrary fiducial hypersurface $t=t_0$. We remark that $a_0=\Gamma_0=1$. The relation between this metric parametrization and the standard metric form and variables of the models is summarized in Appendix B. 

It is useful to describe the dynamics of the models by means of their covariant objects given in terms of the representation of ``q--scalars'' and their perturbations (see \cite{part1,part2} for a comprehensive discussion). For every LTB scalar $A$, the associated q--scalar $A_q$ and perturbation $\Da_q$ are defined by the  correspondence rules 
\footnote{We assume in the integrals in (\ref{Aqdef}) that $r=0$ marks a symmetry center such that $a(t,0),\,\dot a(t,0)$ are nonzero and bounded and $\Gamma(t,0)=1$ holds for all $t$. We also exclude LTB models whose constant $t$ hypersurfaces have two symmetry centers (spherical topology) or no symmetry centers (``wormhole''  topologies). However, these integrals can also be defined for such models (see \cite{sussBR,RadProfs}).  The q--scalars are related to proper volume averages with weight factor $\sqrt{1-\KK_{q0}r^2}$. See \cite{part1} for a comprehensive discussion.} 
\ba A_q =\frac{\int_0^r{A\,a^3\,\Gamma\,\bar r^2\,\dd\bar r}}{\int_0^r{a^3\,\Gamma\,\bar r^2\,\dd\bar r}}=\frac{3r^3\int_0^r{A\,a^3\,\Gamma\,\bar r^2\,\dd\bar r}}{a^3},\label{Aqdef}\\
\Da =\frac{A-A_q}{A_q} = \frac{rA'_q/A_q}{3\Gamma}=\frac{1}{r^3 a^3 A_q}\int_0^r{A'\,\bar r^3\, a^3\dd \bar r},\label{Dadef}
\ea
where the second and third expressions in the right hand side of (\ref{Dadef}), which follow directly by differentiation and integration by parts of (\ref{Aqdef}), allow us to computate $\Da$ in terms of the gradients $A'_q,\,A'$ and the scale factor $\Gamma$.    

The basic LTB covariant scalars are: (i) the rest--mass density $\rho$, (ii) the Hubble scalar $\HH\equiv \theta/3$ (with $\theta=u^a\,_{;a}$) and (iii) the spatial curvature scalar $\KK\equiv \RR/6$ (with $\RR$ the Ricci scalar of surfaces $t=$ const.). In the q--scalar representation these scalars take the forms of exact perturbations~\cite{part1,part2}
\begin{equation}\fl  \rho=\rho_q(1+\Drho),\qquad \HH=\HH_q(1+\Dh),\qquad \KK=\KK_q(1+\DKK),\label{rhoHHKK}\end{equation}
with their associated q--scalars and perturbations given by: 
\ba
\fl  \frac{8\pi}{3}\rho_q =\frac{8\pi}{3}\frac{\rho_{q0}}{a^3}=\frac{\Omega_{q0}\HH_{q0}^2}{a^3}=\Omega_q\HH_q^2,\label{rhoq}
  \\ \fl \KK_q=\frac{\KK_{q0}}{a^2}=\frac{(\Omega_{q0}-1)\HH_{q0}^2}{a^2}=(\Omega_q-1)\HH_q^2,\label{KKq}
  \\ \fl \HH_q^2 =\left(\frac{\dot a}{a}\right)^2=\frac{8\pi}{3}\rho_q-\KK_q=\frac{8\pi\rho_{q0}}{3a^3}-\frac{\KK_{q0}}{a^2}=\HH_{q0}^2\left[\frac{\Omega_{q0}}{a^3}-\frac{\Omega_{q0}-1}{a^2}\right],      \label{HHq}
  \ea
where the q--scalar $\Omega_q$ is defined as  
\begin{equation}\fl \Omega_q \equiv \frac{8\pi\rho_q}{3\HH_q^2}=\frac{\Omega_{q0}}{\Omega_{q0}-(\Omega_{q0}-1)a},\qquad \Omega_q-1=\frac{\KK_q}{\HH_q^2}=\frac{(\Omega_{q0}-1)a}{\Omega_{q0}-(\Omega_{q0}-1)a},\label{Omdef}\end{equation}
and the perturbations satisfy the following scaling laws:
\ba \fl 1+\Drho= \frac{1+\Drho_0}{\Gamma}, \qquad \frac{2}{3}+\DKK = \frac{2/3+\DKK_0}{\Gamma},\label{perts1}\\
\fl 2\Dh= \Omega_q\Drho-(\Omega_q-1)\DKK,\qquad \DOm = (1-\Omega_q)(\Drho-\DKK).\label{perts2}
\ea
Considering that (\ref{HHq}), (\ref{Omdef}) and (\ref{perts2}) provide algebraic constraints linking any two q--scalars and any two of their perturbations, it is evident that any LTB model can be uniquely specified by selecting as free parameters any two of the following four initial value functions: $\rho_{q0},\,\KK_{q0},\,\HH_{q0},\,\Omega_{q0}$ (notice that initial perturbations can always be obtained from (\ref{Dadef}) evaluated at $t=t_0$) 
\footnote{The bang time $\tbb$, which is often taken as another initial value parameter, is expressible in terms of any two of these primary functions, see equation (\ref{tbb}) in the following section.}.  
As shown in \cite{part2}, the models become fully determined in terms of any two q--scalars and their perturbations, which give rise to a covariant and gauge invariant formalism of exact spherical perturbations
\footnote{The perturbations $\Da$ are distinct from the standard metric induced ``gauge invariant'' perturbations \cite{bardeen} and from the covariant perturbations based on the 1+3 formalism \cite{ellisbruni89, BDE, LRS, 1plus3, zibin, dunsbyetal}. Their relation with  simple ``contrast'' perturbations ``$A/A_b-1$'' with respect to an FLRW background value $A_b(t)$ \cite{contrast} is discussed in detail in \cite{part2}.}. 

The main proper tensors of the models, the shear tensor ($\sigma_{ab}=h_a^ch_b^d u_{(c;d)}-\HH\,h_{ab}$) and the electric Weyl tensor ($E_{ab}=u^cu^d C_{acbd}$ with $C_{abcd}$ the Weyl tensor), are expressible in terms of their eigenvalues $\Sigma$ and $\Psi_2$ through a common symmetric trace--free tensor  $\hbox{\bf{e}}^a_b=h^a_b-3n^a n_b$,  with $n_a=\sqrt{g_{rr}}\delta^r_a$ \cite{LRS}:
\ba \fl \sigma^a_b=\Sigma\,\hbox{\bf{e}}^a_b,\qquad \Sigma =\frac{1}{6}\hbox{\bf{e}}_{ab}\sigma^{ab}=-\frac{\dot \Gamma}{3\Gamma}=-(\HH-\HH_q)=-\HH_q\Dh,\label{Sigma1}\\
\fl E^a_b=\Psi_2\,\hbox{\bf{e}}^a_b,\qquad \Psi_2 =\frac{1}{6}\hbox{\bf{e}}_{ab}E^{ab}=\frac{4\pi}{3} (\rho-\rho_q)=\frac{4\pi}{3}\rho_q\Drho,\label{EE1}
\ea
where $\Psi_2$ is also the only nonzero conformal invariant in a Newman--Penrose tetrad representation.    

\section{Analytic solutions as constraints among q--scalars.}

In order to obtain fully determined analytic forms for the q--scalars $A_q$ and the perturbations $\Da$ as functions of time we need the solutions of the Friedman equation (\ref{HHq}), which take the implicit form:  
\begin{equation} \tau_q  \equiv t-\tbb =\tau_q(\rho_q,\KK_q) = \tau_q(\Omega_q,\HH_q),\label{Tq}\end{equation}
where $\tbb=\tbb( r)$ is the Big Bang time such that $a(\tbb,r)=0$ for all $r$. Notice that $\tau_q$ is a function of two q--scalars, hence it is itself a q--scalar \cite{sussbol} having an invariant meaning (from the fact that $\dot\tau_q=1$): it is the total proper time (or affine parameter) length of the worldllines of comoving dust layers. 

The functional forms of $\tau_q=\tau_q(\HH_q,\,\Omega_q)$ are given explicitly below (notice that these expressions yield $\tau_q$ as a function of $a$ by substitution of $\HH_q,\,\Omega_q$ from the scaling laws (\ref{HHq}) and (\ref{Omdef})):  
\footnote{We consider henceforth only hyperbolic and elliptic models or regions, thus we assume that $\Omega_q\ne 1$ and $\KK_q\ne 0$ hold for all $r$. The parabolic case follows as the limit $\Omega_q\to 1$ or $\KK_q\to 0$ and is discussed in section 10.}
\ba
\fl {\underline{\hbox{hyperbolic models:}}}\quad 0<\Omega_q<1,\,(\hbox{or}\,\, \KK_q<0) \nonumber\\  
 \tau_q = \frac{Y_q(\Omega_q)}{\HH_q},\label{hypsol}\\
 \fl {\underline{\hbox{elliptic models:}}}\qquad \Omega_q>1,\,(\hbox{or}\,\,\KK_q>0)\nonumber\\
\fl \tau_q= \left\{ \begin{array}{l}
 Y_q(\Omega_q)/\HH_{q},\qquad\qquad\qquad  
 {\hbox{expanding stage}}\quad \HH_q>0,\\ 
 2\pi\beta_q - Y_q(\Omega_q)/\HH_q, \qquad 
 {\hbox{collapsing stage}}\quad \HH_q<0,\\ 
 \end{array} \right.\label{ellsol}\ea
with $\beta_{q}$ and $Y_q=Y_q(\Omega_q)$ given by
\ba  
 \beta_q = \frac{4\pi\rho_q}{3|\KK_q|^{3/2}}=\frac{\Omega_q}{2|1-\Omega_q|^{3/2}\HH_q}=\beta_{q0}\qquad\left(\,\,\Rightarrow \dot\beta_q=0\right),\label{beta}\\
 Y_q(\Omega_q) = \frac{\epsilon}{|1-\Omega_q|}\left[1-\frac{\Omega_q}{2|1-\Omega_q|^{1/2}}\ACal\left(\frac{2}{\Omega_q}-1\right)\right], \label{Y}
\ea
where $\epsilon =1,\, \ACal=$ arccosh correspond to the hyperbolic case and $\epsilon =-1,\, \ACal=$ arccos to the elliptic case. 

Substitution of $t=t_0$ in (\ref{hypsol}) and the expanding stage of (\ref{ellsol}) yields the Big Bang time and its gradient in terms of initial value functions and perturbations:
\ba \fl \tbb=t_0-\tau_{q0} = t_0-\frac{Y_q(\Omega_{q0})}{\HH_{q0}},\label{tbb}\\
\fl \frac{r}{3}\tbb' = -\tau_{q0}\DT_0 = -\frac{\HH_{q0}\tau_{q0}\Dbe_0+\Dal_0}{\HH_{q0}},\label{DT0}\ea
with 
\bse\ba \fl \Dbe_0=\frac{r}{3}\frac{\beta'_{q0}}{\beta_{q0}}= \Drho_0-\frac{3}{2}\DKK_0= \frac{2+\Omega_{q0}}{2(1-\Omega_{q0})}\DOm_0-\Dh_0,\label{Dbe0}\\
\fl \Dal_0=\frac{r}{3}\frac{\gamma'_{q0}}{\gamma_{q0}}=\DKK_0-\Drho_0=-\frac{\DOm_0}{1-\Omega_{q0}},\quad \left(\gamma_{q0}=\frac{3\KK_{q0}}{4\pi\rho_{q0}}=\frac{2(\Omega_{q0}-1)}{\Omega_{q0}}\right)\label{Dal0}
\ea\ese
For hyperbolic models we have $\tau_q>0$, but for elliptic models $\tau_q$ is restricted by $0<\tau_q<\Tqcoll$, with:
\bse\ba 
\fl \hbox{elliptic expanding:}\quad 0< \tau_q \leq \Tqmax,\qquad \Tqmax = \tmax-\tbb=\pi\beta_{q0},\label{rTqe1}\\
\fl \hbox{elliptic collapsing:}\quad  \Tqmax < \tau_q<\Tqcoll,\qquad \Tqcoll=\tcoll-\tbb=2\pi\beta_{q0},\label{rTqe2}
\ea\ese
where $t=\tmax$ and $t=\tcoll$ mark the times of maximal expansion ($\HH_q=0$) and the collapse singularity (``Big Crunch'' $\HH_q\to-\infty$).   

\section{Growing and decaying density modes.}\label{modes1}

Since the density perturbation $\Drho$ generalizes dust density perturbations in the linear regime, it is worthwhile verifying if it can be decomposed in terms of growing and decaying modes. For this purpose, we use the exact solutions (\ref{hypsol}) and (\ref{ellsol}) to rewrite $\Drho$ in (\ref{perts1}) as 
\begin{figure}
\begin{center}
\includegraphics[scale=0.4]{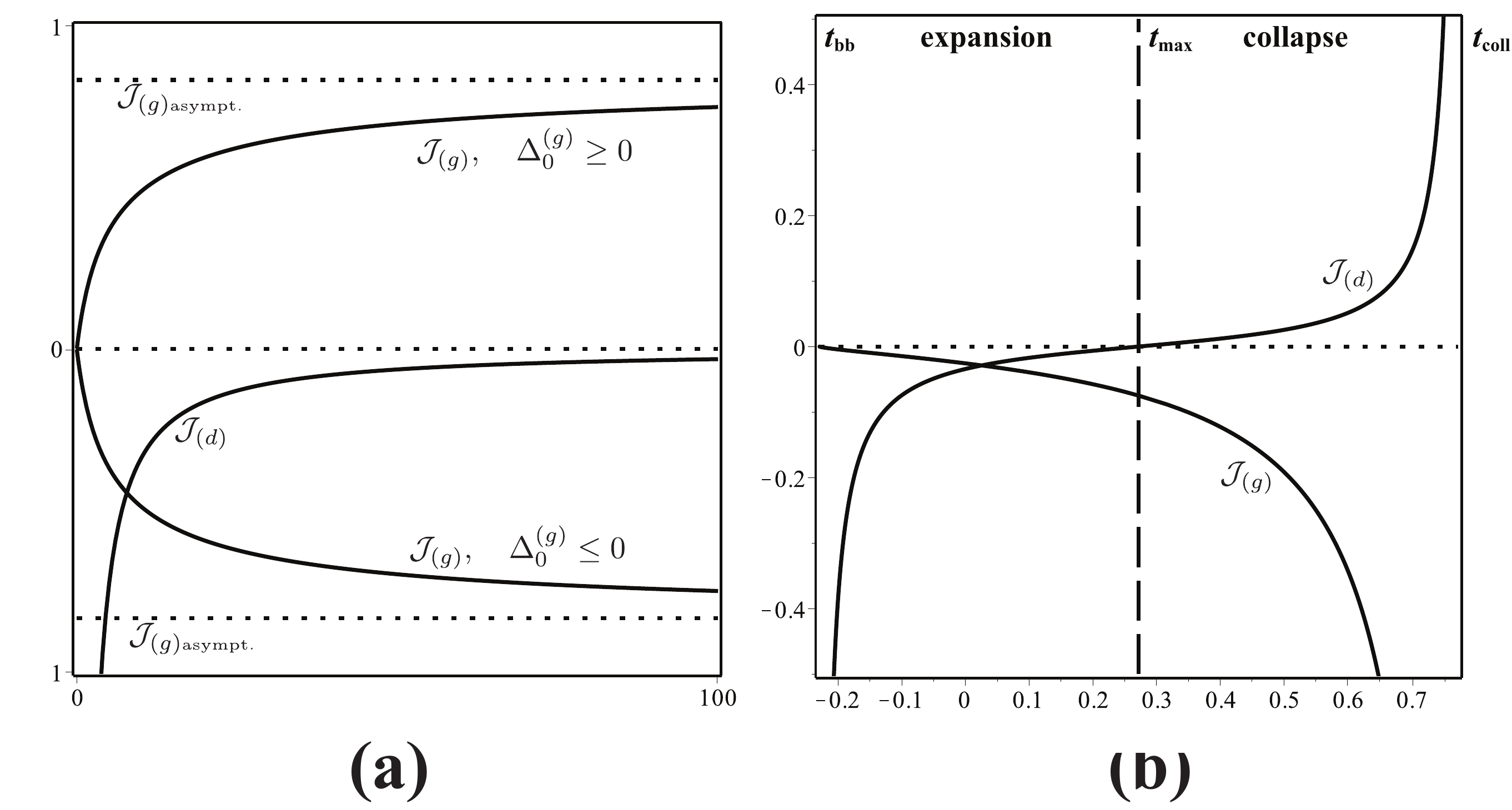}
\caption{{\bf The growing and decaying modes.} The evolution of the density modes, $\Jg$ and $\Jd$, defined by (\ref{gmode})--(\ref{dmode}), is depicted as  functions of time for a typical dust layer. Time units in the horizontal axis are scaled with $1/\HH_{q0}(0)$ for $t_0=0$. Panels (a) and (b), respectively correspond to generic hyperbolic and elliptic models in which both modes are nonzero. Since we assume  compliance with the Hellaby--Lake (HL) conditions to avoid shell crossings (see sections 7--8), we have  $\Did\leq 0$ and $\Jd\leq 0$ in hyperbolic models and the expanding stage of elliptic models (though $\Jd\geq 0$ in the collapsing stage). For hyperbolic models $\Dig$ can be positive or negative, resulting in positive or negative $\Jg$ with asymptotic values $\Jg{}_{\hbox{\tiny{asympt}}}$ given by (\ref{jghas2}), but in elliptic models $\Dig\geq 0$ follows from the HL conditions, and thus $\Jg$ is negative for the whole evolution (see sections 6--12). Notice that $|\Jd|\gg |\Jg|$ and $|\Jg|\gg|\Jd|$ respectively hold in the early and late time evolution.}
\label{fig1}
\end{center}
\end{figure} 
\ba \Drho = \frac{1+\Drho_0-\Gamma}{\Gamma}= \frac{\Jg+\Jd}{1-\Jg-\Jd},\label{DrhoJ} \\
\Gamma = (1+\Drho_0)(1-\Jg-\Jd),\label{GammaJ}\ea
in which we identify:
\ba \Jg = 3\Dig\left(\HH_q \tau_q-\frac{2}{3}\right),\qquad \hbox{density growing mode},\label{gmode}\\
\Jd = 3\Did\,\HH_q,\qquad\qquad\qquad \hbox{density decaying mode}, \label{dmode}\ea
with the coefficients or ``amplitudes'' of the modes, $\Dig$ and $\Did$ (both assumed nonzero unless stated otherwise)  given in terms of primary initial value functions as
\ba \Dig \equiv \frac{\Dbe_0}{1+\Drho_0},\qquad \Did \equiv -\frac{\tau_{q0}\DT_0}{1+\Drho_0}=\frac{r\tbb'}{3(1+\Drho_0)},\label{amplitudes}\ea
with $\tau_{q0}\DT_0$ and $\Dbe_0$ given by (\ref{DT0}) and (\ref{Dbe0}). The scalars $\HH_q\tau_q-2/3$ and $\HH_q$, together with the restrictions on $\Dig$ and $\Did$ from the conditions to avoid shell crossings (see sections 7--8), determine the time evolution of $\Jg$ and $\Jd$ that is displayed in figure 1. We provide in Appendix A a review and comparison with exact expressions obtained for these modes in previous literature \cite{kras2,wainwright,silk}, while Appendix B illustrates how these expressions can be computed in the traditional variables.

\subsection{Properties of the density modes.}

\subsubsection{The modes amplitudes are fluid conserved quantities.} It is straightforward to prove by means of (\ref{perts1})--(\ref{perts2}) and (\ref{DT0})--(\ref{Dal0}) that the amplitudes $\Dig$ and $\Did$ in (\ref{amplitudes}) are fluid conserved quantities:
\ba \fl \Dg = \frac{\Dbe}{1+\Drho}=\Dig,\quad \Dd =-\frac{\tau_q\DT}{1+\Drho}=\Did \quad \Rightarrow\quad \dot \Dg=\dot \Dd=0,\label{DgDd}\ea
which follows from $1+\Drho_0=(1+\Drho)\Gamma$ in (\ref{perts1}) and the scaling laws
\begin{equation}\Dbe_0=\Dbe\,\Gamma,\qquad \tau_{q0}\DT_0 =\tau_q\DT\,\Gamma,\label{idents}\end{equation}
where $\Dbe$ and $\tau_q\DT$ are the general forms for arbitrary $t\ne t_0$ of $\Dbe_0$ and $\tau_q\DT_0$ given by (\ref{DT0}) and (\ref{Dbe0}). The fluid conservation of their amplitudes allows us to write the growing and decaying modes as constraints between q-scalars and perturbations whose forms are also preserved by the fluid flow:
\begin{equation}\Jg=3\Dg\left(\HH_q\tau_q-\frac{2}{3}\right),\qquad \Jd =3\Dd\HH_q,\label{modes}\end{equation}
with the conserved amplitudes $\Dg$ and $\Dd$ given by (\ref{DgDd}). 

\subsubsection{The density modes are coordinate independent quantities.} This follows from the fact that $\Jg$ and $\Jd$  are constructed with $\HH_q=\HH+\Sigma$ and $\tau_q=t-\tbb$ (total proper time length along integral curves of the 4--velocity), while the amplitudes $\Dig=\Dg$ and $\Did=\Dd$ are fluid conserved quantities given by (\ref{DgDd}).

\subsubsection{The growing mode and spatial curvature.}  The sign of the dimensionless covariant q--scalar $\HH_q\tau_q-2/3$ in the growing mode $\Jg$ in (\ref{gmode}) is closely related to the sign of the q--scalar associated to the spatial curvature $\KK_q$ (or $\Omega_q-1$, see (\ref{Omdef})):
\footnote{Notice that $\KK_q$ is not the spatial curvature $\KK$ defined  in (\ref{rhoHHKK}). While the sign of $\KK$ determines the sign of $\KK_q$, the converse is false: examples of elliptic LTB models (for which $\KK_q>0$ holds everywhere) exist in which $\KK<0$ holds in some domains (see \cite{sussBR}).}
\begin{equation}\fl  \HH_q(t-\tbb)-\frac{2}{3} \quad  \left\{ \begin{array}{l}
 >0 \quad \Leftrightarrow\quad \Omega_1<1\,\,\hbox{or}\,\,\KK_q<0,\quad\hbox{hyperbolic layers},\\ 
 =0 \quad \Leftrightarrow\quad \Omega_1=1\,\,\hbox{or}\,\,\KK_q=0,\quad\hbox{parabolic layers},\\
 <0 \quad \Leftrightarrow\quad \Omega_1>1\,\,\hbox{or}\,\,\KK_q>0,\quad\hbox{elliptic layers},
 \end{array} \right.\label{Hqt23}\end{equation}
and thus provides a relation between $\Jg$ and the type of kinematic evolution of dust layers as determined by the spatial curvature through $\KK_q$ or $\Omega_q-1$. Since these q--scalars are the analogues of the spatial curvature in FLRW models, the sign relation (\ref{Hqt23}) provides an exact covariant generalization to the relation between the growing mode and the deviation from spatial flatness in linear perturbations around an Einstein--de Sitter background (see Appendix of \cite{zibin}). We comment on this issue in the following section. 

\subsubsection{The remaining perturbations.} Considering the scaling laws (\ref{perts1})--(\ref{perts2}) and the form of $\Drho$ and $\Gamma$ in  (\ref{DrhoJ})--(\ref{GammaJ}), we can express the remaining  perturbations in terms of the growing/decaying modes $\Jg,\,\Jd$:  
\bse\ba
  \fl\DKK &=& \frac{2\,(\Jg+\Jd-\Dig)}{3(1-\Jg-\Jd)}=\frac{2\,\left[\Dig(\HH_q\tau_q-1)+\Did\HH_q\right]}{1-\Jg-\Jd},\label{DK}\\
 \fl \DOm &=& \frac{(1-\Omega_q)\,\left(\Jg+\Jd+2\Dig\right)}{3(1-\Jg-\Jd)}=\frac{(1-\Omega_q)\,\left[\Dig\HH_q\tau_q+\Did\HH_q\right]}{1-\Jg-\Jd},\label{DOM}\\
 \fl \Dh &=& \frac{(2+\Omega_q)(\Jg+\Jd)-2(1-\Omega_q)\Dig}{6(1-\Jg-\Jd)}\nonumber\\
 \fl &=&\frac{\Dig\left[\HH_q\tau_q(2+\Omega_q)-2\right]+\Did (2+\Omega_q)\HH_q}{2(1-\Jg-\Jd)},\label{DH}\ea\ese
where the following forms of $\HH_q\tau_q$ and $\HH_q$ are useful for the purpose of computations:
\bse\ba
\fl \HH_q\tau_q=Y_q,\qquad \hbox{(hyperbolic \& elliptic expanding)},
\label{HTq}\\
\fl \HH_q\tau_q=Y_q-\frac{\pi\Omega_q}{(\Omega_q-1)^{3/2}},\quad\hbox{(elliptic collapsing)}, \label{HTqc}\ea\ese
\begin{equation}\fl \HH_q =\pm\HH_{q0}\frac{\Omega_q}{\Omega_{q0}}\,\left|\frac{1-\Omega_{q0}}{1-\Omega_q}\right|^{3/2},\label{HB}\end{equation}
where $Y_q=Y_q(\Omega_q)$ is given by (\ref{Y}) and we have assumed that $\HH_{q0}>0$. Notice that both $\HH_q \tau_q$ and $\HH_q$ above can also be expressed in terms of the scale factor $a$ by means of the scaling laws (\ref{HHq}) and (\ref{Omdef}). 

\section{Linear limit.}

The exact form (\ref{DrhoJ}) illustrates the expected non--linear dependence of the density perturbation $\Drho$ on the coupled modes $\Jg$ and $\Jd$, as $\Drho$ is not a linear perturbation, but an exact perturbation. In fact, $\Drho$ is a solution of the non--linear evolution equation \cite{part2}
\begin{equation} \ddot\Drho-\frac{2[\dot\Drho]^2}{1+\Drho}+2\HH_q\dot\Drho-4\pi\rho_q\Drho(1+\Drho)=0.\label{nlineq}\end{equation}
However, under linear conditions characterized by $|\Dig|\ll 1$ and $|\Did|\ll 1$ we recover well know results of the linear theory. A series expansion of (\ref{DrhoJ}) around $\Dig$ and $\Did$ up to leading terms yields $\Drho$ as a linear combination of the growing and decaying mode: 
\begin{equation} \Drho \approx \Jg+\Jd,\label{Dmlin}\end{equation}
with $\Jg,\,\Jd$ given by (\ref{gmode}) and (\ref{dmode}), which is the expected result of linear theory in which the growth of $\Drho$ is directly controlled by the interplay between the completely decoupled modes $\Jg,\,\Jd$, and thus it justifies the non--linear exact form (\ref{DrhoJ}) in which both modes are necessarily coupled. Notice that in the linear regime (\ref{Dmlin}) the density perturbation diverges as $t\to\tbb$ if $\Did\ne 0$ (because $\HH_q\to\infty$ in (\ref{dmode})), while it remains finite in the non--linear regime (\ref{DrhoJ}) (see (\ref{Dma0}) further ahead).  

In the linear regime we have $\rho_q\approx \rho$ and $\HH_q\approx \HH$, and thus (\ref{Dmlin}) is the solution of the following equation for $\Drho$ furnished by the linear limit of (\ref{nlineq}):
\begin{equation} \ddot\Drho+2\HH\dot\Drho-4\pi\rho\,\Drho=0.\label{lineq}\end{equation}
which is the known evolution equation for linear dust perturbations in the comoving gauge \cite{part2,bardeen,contrast} once we consider that $\rho$ and $\HH$ are close to their FLRW values $\rho_{\tiny{\textrm{FLRW}}}$ and $\HH_{\tiny{\textrm{FLRW}}}$. 

The particular case of (\ref{lineq}) for a spatially flat FLRW background illustrates in a striking manner the direct relation between the density modes and the growth of $\Drho$
\footnote{The growth of the ``density contrast'' is often computed with the simple construction $\rho/\rho_b(t)-1$, where $\rho$ is the local density of an inhomogeneous model and $\rho_b(t)$ is the density of a suitable FLRW background. As shown in \cite{part2}, these simple ``contrast perturbations'' follow as the asymptotic limit $\rho_q\to\rho_b(t)$ as $r\to\infty$ of non--local perturbations defined for LTB models that converge asymptotically to a FLRW state in the radial direction \cite{RadAs}. While $\Drho$ and the ``contrast perturbation" are different objects, they approximate each other in the linear limit.  See \cite{part2} for a comprehensive discussion. }
. Expanding the solutions (\ref{hypsol})--(\ref{ellsol}), as well as $\Jg$ and $\Jd$ given by (\ref{gmode}) and (\ref{dmode}), around $\Omega_q\approx 1$ (or $\KK_q\approx 0$) yields 
\bse\ba \Jg \approx  -\frac{2}{5}\Dig(\Omega_q-1)\approx  -\frac{2}{5}\,\frac{(\Omega_{q0}-1)\Dig}{\Omega_{q0}}\,a,\label{lingmode}\\
\Jd\approx  \frac{3\Did |1-\Omega_{q0}|^{3/2}}{\Omega_{q0}|1-\Omega_q|^{3/2}}\approx -\frac{3\sqrt{\Omega_{q0}}\,\Did}{a^{3/2}},\label{lindmode}\\
\tau_q \approx -\frac{2}{3}\Omega_{q0}\frac{|\Omega_q-1|^{3/2}}{|\Omega_{q0}-1|^{3/2}}\approx -\frac{2}{3}\frac{a^{3/2}}{\sqrt{\Omega_{q0}}}, \label{tqlin}\ea\ese 
where we used (\ref{HTq}) and (\ref{HB}), and then (\ref{Omdef}) to express the terms containing $\Omega_q$ in terms of $a$. It is worthwhile comparing these expansions with those obtained by Zibin \cite{zibin} in looking at linear perturbations on an Einstein de Sitter background in the context of LTB models. From its form in (\ref{Dmlin}) and considering (\ref{lingmode})--(\ref{tqlin}), the linear limit of $\Drho$ becomes formally identical to Zibin's equation (A1) in Appendix of \cite{zibin}, which is the familiar expression of the linear dust density perturbation found in the literature (see also \cite{contrast}). The expansion of $\Jg$ in (\ref{lingmode}), which coincides with Zibin's equation (A3) in \cite{zibin}, illustrates the relation between the growing mode and the small deviations from spatial flatness ({\it i.e.} what Zibin in \cite{zibin} calls the spatial curvature ``fluctuations'') of linear perturbations around an an Einstein--de Sitter background in which $|\Omega_q-1|$ is expected to be very small. This relation between $\Jg$ and $\Omega_q-1$ is described by Zibin as showing that ``the curvature perturbation consists of {\it just the growing mode}''. However, this statement is not consistent with the quasi--local perturbation formalism described in \cite{part2}. The linear limits of the remaining exact perturbations in (\ref{DK})--(\ref{DH}): 
\ba \fl \Dh \approx \frac{1}{3}\Drho,\quad \DKK\approx -\frac{\Dig}{3}-\frac{1}{3}\Jd,\quad \DOm\approx -\frac{5}{3}\Jg-\frac{\Omega_{q0}-1}{3\Omega_{q0}^{3/2}}\,a\,\Jd.\label{linD}\ea 
show that the spatial curvature perturbation $\DKK$ also depends on the decaying mode in its linar limit. Rather, in the framework of this formalism, the relation between $\Jg$ and $\Omega_q-1$ in (\ref{lingmode}) is simply the perturbative version (around $\Omega_q=1$) of the exact relation (\ref{Hqt23}) connecting the growing mode and spatial curvature. Since spatial flatness ($\Omega_q-1=0$) yields exactly $\HH_q-2/3=0$, linear dust perturbations around an Einstein--de Sitter background will necessarily yield a perturbative form of the growing mode (\ref{gmode}) that is defined by small deviations from spatial flatness, but this does not imply that the spatial curvature perturbation is only defined by the growing mode. Notice that (\ref{Hqt23}) also holds in the linear limit where $\HH_q\approx \HH\approx \HH_{\tiny{\textrm{FLRW}}},\,\KK_q\approx \KK\approx \KK_{\tiny{\textrm{FLRW}}}$ and $\tbb=0$, taking the form $\HH_{\tiny{\textrm{FLRW}}}\,t-2/3$, with its sign following the same relation with the curvature of FLRW time slices as (\ref{Hqt23}) does with the quasi--local curvature of LTB time slices.   

\begin{table}
\begin{center}
\begin{tabular}{|c| c| c|}
\hline
\hline
\hline
\hline
\hline
\hline
\multicolumn{3}{|c|}{3-dimensional regions}
\\  
\hline
\hline
{Name} &{Description} &{Phase Space constraints} 
\\
\hline
\hline
{{\bf HYP}} &{Hyperbolic models} &{$0<\Omega_q<1$}
\\  
{} &{of the general case} &{$\jg+\jd+1\geq 0$}
\\  
{} &{complying with} &{$1+\jd-\frac{(1-W_q)}{W_q-2/3}\jg\geq 0$} 
\\  
{} &{Hellaby-Lake conditions.} &{$\Dd\leq  0\;\;\Rightarrow\;\;\jd\leq 0$.}
\\  
\hline
\hline
{{\bf ELL}} &{Elliptic models} &{$\Omega_q>1$}
\\  
{} &{of the general case} &{$\jg+\jd+1\geq 0$}
\\  
{} &{complying with} &{$\jd-\frac{\pi\Omega_q}{(\Omega_q-1)^{3/2}(\frac{2}{3}-W_q)}\jg\geq 0$} 
\\  
{} &{Hellaby-Lake conditions.} &{$\Dg\geq 0\;\;\Rightarrow\;\; \jg\leq 0$}
\\  
{} &{} &{$\Dd\leq  0\;\;\Rightarrow\;\;\jd\leq 0$\;\;($\HH_q>0$).}
\\
{} &{} &{$\Dd\leq  0\;\;\Rightarrow\;\;\jd\geq 0$\;\;($\HH_q<0$).}
\\
\hline
\hline  
\hline
\hline
\hline
\hline
\multicolumn{3}{|c|}{2--dimensional planes}
\\  
\hline
\hline
{Name} &{Description} &{Phase Space constraints} 
\\
\hline
\hline
{{\bf SDW}} &{Suppressed Decaying Mode} &{$\jd =0, \quad[\jg,\Omega_q]$} 
\\
{} &{(simultaneous big bang).} &{$W_q\ne 2/3,\quad \Omega_q\ne 1$.} 
\\
\hline
{{\bf SGW}} &{Suppressed Growing Mode} &{$\jg =0, \quad[\jd,\Omega_q]$} 
\\
{} &{elliptic \& hyperbolic.} &{$W_q\ne 2/3,\quad \Omega_q\ne 1$.} 
\\
\hline
\hline
{{\bf VAC}} &{Minkowski vacuum in} &{$\Omega_q=0,\quad[\jd,\,\jg].$}
\\
{} &{non--standard coordinates. } &{} 
\\
\hline 
\hline
\hline
\hline
\hline
\hline
\multicolumn{3}{|c|}{Lines and points. }
\\  
\hline
\hline
{Name} &{Description} &{Phase Space constraints} 
\\ 
\hline
\hline
{{\bf PAR}} &{Parabolic Models} &{$\jg=0,\quad\jd=\jd(\xi)$} 
\\
{} &{subset of {\bf SGM}.} &{$W_q= 2/3,\quad\Omega_q=1$.} 
\\
\hline
{{\bf FLRW}} &{FLRW dust models} &{$\jg=\jd=0,\quad\Omega_q=\Omega_q(\xi)$.}
\\
{} &{\& center worldline.} &{} 
\\
\hline
{{\bf EDS}} &{Einstein De Sitter Model} &{$\jg=\jd=0,\quad \Omega_q=1$.} 
\\ 
{} &{subset of {\bf{FLRW}}.} &{} 
\\
\hline
{{\bf MINK}} &{Minkowski} &{$\jg=\jd=\Omega_q=0$.} 
\\ 
{} &{subset of {\bf{VAC}}.} &{}
\\
\hline
\hline
\end{tabular}
\end{center}
\caption{{\bf{Invariant subspaces of the phase space $\PP$}}. The function $W_q$ is defined by (\ref{Wqdef}). All non--trivial LTB models can be classified in terms of their evolution in subclasses defined by the invariant subspaces {\bf HYP}, {\bf ELL}, {\bf SDM}, {\bf SGM}, {\bf PAR} and {\bf VAC}. Notice that {\bf VAC} corresponds to the particular LTB vacuum solution of the Friedman equation (\ref{HHq}) with $\rho_{q0}=\Omega_{q0}=0$,\, $\KK_{q0}<0$ and $\HH_{q0}=\sqrt{|\KK_{q0}|}$ (see \cite{ltbstuff,sussBR,RadAs}). Phase space evolution of dust layers in these invariant subspaces is examined in sections 7--10 and critical points are listed in Tables 2, 3 and 4. The Hellaby--Lake conditions to avoid shell crossings are discussed in sections 7 and 8.}
\label{tabla1}
\end{table}

\section{The growing/decaying modes vs phase space invariant subspaces.}

A phase space description is the ideal tool to examine the role of the growing and decaying modes in the dynamics of the models (see \cite{sussDS1,sussDS2} for previous work on dynamical systems applied to LTB models). For this purpose, it is convenient to construct a dynamical system whose phase space is parametrized by the following bounded variables directly related to $\Jg$ and $\Jd$ as defined in (\ref{modes}):
\footnote{These variables are ill defined if there is a shell crossing singularity in which $\Gamma=0$ occurs at $a>0$ (or equivalently $\Omega_q>0$). }
\bse\ba \fl \jg \equiv \frac{\Jg}{1-\Jg-\Jd}= (1+\Drho)\Jg=3\Dbe\left(W_q-\frac{2}{3}\right)=\frac{3\Dbe_0}{\Gamma}\left(W_q-\frac{2}{3}\right),\label{jg}\\
\fl \jd \equiv \frac{\Jd}{1-\Jg-\Jd}= (1+\Drho)\Jd=-3\tau_q\DT\HH_q=-\frac{3\tau_{q0}\DT_0}{\Gamma}\HH_q,\label{jd}\ea\ese
where we used (\ref{idents}), with $\Dbe$ and $\tau_q\DT$ given in terms of primary perturbations by the generalization for $t\ne t_0$ of (\ref{DT0}) and (\ref{Dbe0}), while $W_q\equiv \HH_q\tau_q$ takes the form:
\ba\fl W_q= \left\{ \begin{array}{l}
 Y_q\,\,\hbox{if}\,\,\HH_q>0,\quad  
 {\hbox{(hyperbolic \& elliptic expanding)}},\\ 
 Y_q-\pi\Omega_q/(\Omega_q-1)^{3/2}\,\,\hbox{if}\,\,\HH_q<0,\quad 
 {\hbox{(elliptic collapsing)}},\\ 
 \end{array} \right.
 \label{Wqdef}\ea
in which we considered all possible forms of $\HH_q\tau_q$ in (\ref{HTq})--(\ref{HTqc}) with $Y_q=Y_q(\Omega_q)$ defined by (\ref{Y}).  

The scaling laws for the q--scalars and their perturbations that have been previously derived [see  (\ref{rhoq})--(\ref{perts2}), (\ref{DrhoJ}), (\ref{GammaJ}), (\ref{DK})--(\ref{DH})] are analytic solutions of the evolution equations for the models, such as the following ones in the representation $\{\HH_q,\,\Omega_q,\,\Dh,\,\DOm\}$ \cite{part2}:  
\bse\ba \dot\HH_q &=& -(1+\frac{1}{2}\Omega_q)\,\HH_q^2,\label{EV1}\\
\dot \Omega_q &=& -\Omega_q(1-\Omega_q)\,\HH_q,\label{EV2}\\
\dot \Dh &=& -\left[(1+3\Dh)\Dh+\frac{1}{2}\Omega_q(\Dh+\DOm)\right]\,\HH_q,\label{EV3}\\
\dot \DOm &=& -\left[(1+3\DOm)\Dh-\Omega_q(\Dh+\DOm)\right]\,\HH_q.\label{EV4}\ea\ese
in which the expanding ($\HH_q>0$) and collapsing ($\HH_q<0$) stages of elliptic models must be treated separately, since $\Omega_q,\,\DOm$ and $\Dh$ diverge as $\HH_q\to 0,\,t\to\tmax$. 

By eliminating $\DOm$ and $\Dh$ in terms of $\jg$ and $\jd$ from (\ref{Dbe0})--(\ref{DT0}) and (\ref{jg})--(\ref{jd}):
\bse\ba \Dh=\frac{[(2+\Omega_q)W_q-2]\jg+(2+\Omega_q)(W_q-\frac{2}{3})\jd}{6(W_q-\frac{2}{3})},\label{Dh2}\\
  \DOm=\frac{(1-\Omega_q)[W_q\jg+(W_q-\frac{2}{3})\jd]}{3(W_q-\frac{2}{3})},\label{DOm2}\ea\ese 
and then substituting in (\ref{EV1})--(\ref{EV4}) we obtain the following self consistent dynamical system that is valid for all LTB models: 
\bse\ba \fl \frac{\partial\Omega_q}{\partial\xi} &=& \varepsilon\,\Omega_q(\Omega_q-1),\label{DS1}\\
\fl\frac{\partial\jg}{\partial\xi} &=& \varepsilon\,\frac{[2-(2+\Omega_q)W_q]\jg^2}{2(W_q-2/3)}-\varepsilon\,\left[\frac{(2+\Omega_q)\jd}{2}-\frac{\Omega_q(\Omega_q-1)}{W_q-2/3}\frac{\dd W_q}{\dd\Omega_q}\right]\jg, \label{DS2}\\
\fl \frac{\partial\jd}{\partial\xi} &=& \varepsilon\,\left[\frac{[2-(2+\Omega_q)W_q](1+\jd)}{2(W_q-2/3)}-\frac{\Omega_q(\Omega_q-1)}{W_q-2/3}\frac{\dd W_q}{\dd\Omega_q}\right]\jg\nonumber\\
\fl &-&\varepsilon\,\frac{(2+\Omega_q)\jd(1+\jd)}{2},\label{DS3}
\ea\ese
where $W_q$ is defined by (\ref{Wqdef}) and  $\varepsilon=1,\,-1$ respectively correspond to $\HH_q>0$ (hyperbolic \& elliptic expanding) and $\HH_q<0$ (elliptic collapsing), and the evolution parameter $\xi$ follows from applying the coordinate transformation~\cite{sussDS1,sussDS2}
\begin{equation}\fl \xi =\xi(t,r),\quad \bar r = r,\qquad \frac{1}{\HH_q}\left(\frac{\partial}{\partial t}\right)_r=\left(\frac{\partial}{\partial \xi}\right)_r,\qquad \xi =\varepsilon\,\ln\, a,\label{xidef}\end{equation}
so that $\xi=0$ corresponds to $a=1$ that defines the initial slice $t=t_0$ (though the rest of the surfaces $\xi$ constant do not coincide with $t$ constant surfaces for $t\ne t_0$). 

The evolution of each dust layer ($r$ constant) in any given LTB model becomes a trajectory $\UU(\xi,r)=[\jg(\xi,r),\,\jd(\xi,r),\,\Omega_q(\xi,r)]$ in the 3--dimensional phase space $\PP=[\jg,\,\jd,\,\Omega_q]$ associated with (\ref{DS1})--(\ref{DS3}), determined by initial conditions specified at $\xi=0$
\bse\ba \fl \UU(0,r)=[\Omega_{q0},\,\jg{}_0,\,\jd{}_0],\label{initconds1}\\ 
\fl \jg{}_0=\jg(0,r)=3\Dbe_0(W_{q0}-2/3),\qquad \jd{}_0=\jd(0,r)=-3\tau_{q0}\DT_0\HH_{q0},\label{initconds2}\ea\ese
where we used the fact that $a=\Gamma=1$ at $\xi=0$ to evaluate the initial values in (\ref{jg})--(\ref{jd}) and $W_{q0}=W_q(\Omega_{q0})$ with $W_q$ given by (\ref{Wqdef}). Evidently, the phase space evolution of each LTB model determines a unique surface $\UU(\xi,r)$ of $\PP$ generated by the trajectories once we consider the full range of $r$ in the initial conditions (\ref{initconds1})--(\ref{initconds2}). 

While the dynamical system (\ref{DS1})--(\ref{DS3}) can be solved numerically, its analytic solutions for any set of initial conditions (\ref{initconds1})--(\ref{initconds2}) are readily available from the expressions previously derived for the phase space variables: $\Omega_q(\xi,r)$ is given  the scaling law (\ref{Omdef}) expressed in terms of $\xi$ in (\ref{xidef}): 
\begin{equation}\fl \Omega_q(\xi,r)=\frac{\Omega_{q0}}{\Omega_{q0}-(\Omega_{q0}-1)\,{\textrm{exp}}(\varepsilon\,\xi)},\qquad \varepsilon=\left\{ \begin{array}{l}
 1,\,\,\HH_q>0,\\ 
 -1,\,\,\HH_q<0,\\ 
 \end{array} \right.\label{Omxi} \end{equation}
which substituted into (\ref{jg})--(\ref{jd}) yields $\jg$ and $\jd$ as explicit functions of $(\xi,r)$ after using (\ref{Y}) to express $Y_q$ in terms of $\Omega_q$ given above. 

The invariant subspaces of the phase space $\PP$ are constraints among the phase space variables $[\Omega_q,\,\jg,\,\jd]$ that are preserved by the fluid flow ({\it i.e.} hold throughout the full time evolution). We remark that all constraints among the q--scalars and the perturbations can always be expressed as constraints among the phase space variables by elimination of $\Drho,\,\DKK$ by means of (\ref{perts1})--(\ref{perts2}), (\ref{DT0}), (\ref{Dbe0}), (\ref{DrhoJ}), (\ref{DK})--(\ref{DH}) and (\ref{jg})--(\ref{jd}). The invariant subspaces of $\PP$ relevant to the dynamics of the models provide a useful classification of subclasses of models and are listed and classified in Table 1, which also provides the conditions for absence of shell crossings given in terms of the phase space coordinates.  In the following sections we examine qualitatively the phase space dynamics of the models, dealing first with the general case in which $\jg$ and $\jd$ are both nonzero, and then with the cases when each one of the density modes is suppressed.
\begin{table}
\begin{center}
\begin{tabular}{|c| c| c|}
\hline
\hline
\hline
\hline
\multicolumn{3}{|c|}{{\bf Critical Points. General case:} $\jg\ne 0,\,\jd\ne 0$.}
\\
\hline
\hline
\hline
\hline
\multicolumn{3}{|c|}{Hyperbolic models. Section 7. Figure 2.}
\\  
\hline
\hline
\hline
\hline
{Symbol} &{Phase Space coordinates} &{Description} 
\\
\hline
\hline
{{\bf BB}} &{$\Omega_q=1,\,\jd=-1,\,\jg=0$} &{Non--simultaneous Big Bang.}
\\  
{} &{} &{Past attractor (source)}
\\
\hline
\hline  
{{\bf MIL}} &{$\Omega_q=0,\,\jd=0,\,\jg=\frac{\Dig}{1-\Dig}$} &{Milne. Future attractors (sinks)} 
\\
\hline
\hline  
{{\bf EdS}} &{$\Omega_q=1,\,\jd=0,\,\jg=0$} &{Einstein--de Sitter. Saddle}
\\  
\hline
\hline
{{\bf S1}} &{$\Omega_q=0,\,\jd=-1,\,\jg=0$} &{Saddle}
\\  
\hline
\hline
{{\bf S2}} &{$\Omega_q=1,\,\jd=0,\,\jg=-1$} &{Saddle}
\\ 
\hline
\hline
\hline
\hline
\multicolumn{3}{|c|}{Elliptic models. Section 8. Figure 3.}
\\
\hline
\hline
\hline
\hline
{Symbol} &{Phase Space coordinates} &{Description} 
\\
\hline
\hline
{{\bf BB}} &{$\Omega_q=1,\,\jd=-1,\,\jg=0$} &{Non--simultaneous Big Bang.}
\\  
{} &{} &{Past attractor (source)}
\\
\hline  
{{\bf BC}} &{$\Omega_q=1,\,\jd=-\jg-1$} &{Big Crunch. } 
\\  
{} &{$\jg=\frac{-\pi \Dig \Omega_{q0}}{\Did(\Omega_{q0}-1)^{3/2}\HH_{q0}+\pi\Dig\Omega_{q0}}$} &{Future attractors (sinks)}
\\  
{} &{} &{}
\\
\hline  
{{\bf EdS}} &{$\Omega_q=1,\,\jd=0,\,\jg=0$} &{Einstein--de Sitter. Saddle}
\\    
\hline
{{\bf S2}} &{$\Omega_q=1,\,\jd=0,\,\jg=-1$} &{Saddle}
\\    
\hline
\hline
\hline
\hline
\end{tabular}
\end{center}
\caption{{\bf{Critical points of $\PP$ for the general case when both density modes are nonzero}}. These critical points (which are displayed by figures 2 and 3) follow from the dynamical system (\ref{DS1})--(\ref{DS3}), with $\varepsilon =1$ for hyperbolic models and elliptic models in their expanding phase ($\HH_q>0$) and $\varepsilon =-1$ for the collapsing stage ($\HH_q<0$) of the latter models. }
\label{tabla2}
\end{table}

\section{The general case: hyperbolic models.}
\begin{figure}
\begin{center}
\includegraphics[scale=0.5]{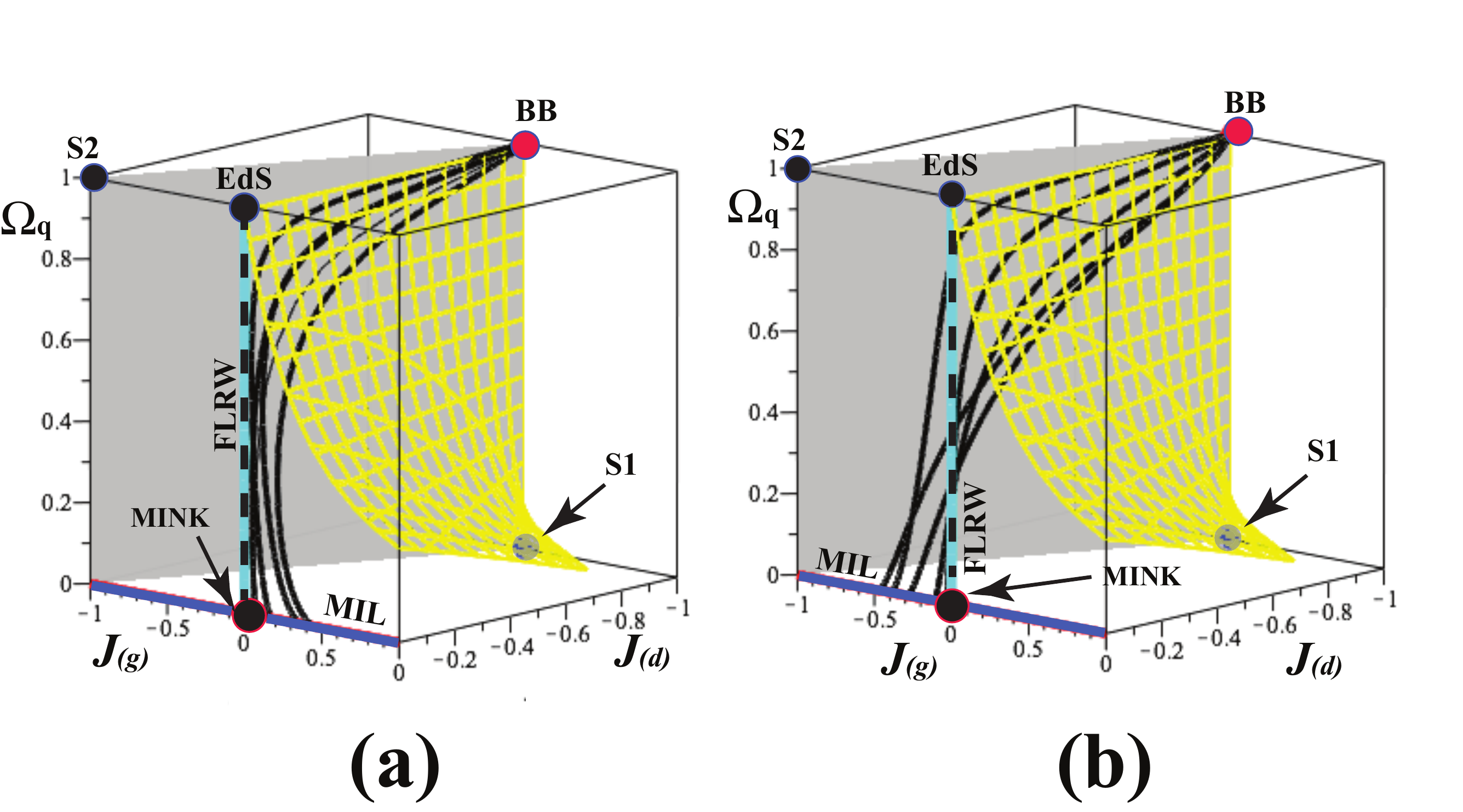}
\caption{{\bf Phase space evolution of the general case: hyperbolic models.} Panels (a) and (b) respectively correspond to the cases when $\Dig\geq 0$ and $\Dig\leq0$. The coordinates of the critical points (bold letters) are given in Table 2. The evolution of phase space trajectories and description of critical points and invariant subspaces are discussed in section 7.4. All trajectories are fully contained inside of the shaded and wireframe surfaces (grey and yellow in the online version) marking the boundary of HL constraints (\ref{shxh2a})--(\ref{shxh2c}) that define the invariant subspace {\bf HYP}.}
\label{fig1}
\end{center}
\end{figure}
\subsection{Absence of shell crossings.}

Phase space trajectories are confined to the invariant subspace $0<\Omega_q<1$. However, avoidance of shell crossing singularities ($\Gamma>0$ holds for all $\tau_q>0$ and all $r$) implies further restrictions on the region of $\PP$ containing the trajectories, as initial conditions must comply with the following constraints (the Hellaby--Lake (HL) conditions) \cite{ltbstuff,RadAs,RadProfs}:
\ba \fl 1+\Drho_0\geq 0, \quad\frac{2}{3}+\DKK_0=\frac{2}{3}(1+\Drho_0)(1-\Dig)\geq 0\quad \hbox{and}\quad \tbb'\leq 0\;\;(\Rightarrow\;\; \Did\leq 0),\nonumber\\
\fl\label{shxh} \ea
whose fulfillment implies the following fluid flow preserved constraints:
\bse\ba \fl 1+\Drho\geq 0 \quad\Rightarrow\quad 1+\jg+\jd\geq 0,\label{shxh2a}\\
\fl \frac{2}{3}+\DKK\geq 0 \quad\Rightarrow\quad \Dg\leq 1 \quad\Rightarrow\quad 1+\jd-\frac{1-Y_q}{Y_q-2/3}\jg\geq 0,\label{shxh2b}\\
\fl \Dd\leq 0 \quad \Rightarrow\quad \jd\leq 0,\label{shxh2c}\ea\ese
where we have taken into consideration that $\HH_q\tau_q=W_q=Y_q,\,\Gamma>0,\,\HH_q>0$ and $Y_q-2/3> 0$ hold, together with using (\ref{DrhoJ}), (\ref{DK}), (\ref{jg})--(\ref{jd}) and (\ref{Dh2})--(\ref{DOm2}) to eliminate $\Drho,\,\DKK$ in terms of $\jg,\,\jd$. Since each one of the HL constraints (\ref{shxh2a})--(\ref{shxh2c}) defines an invariant subspace of $\PP$, the phase space trajectories of all regular hyperbolic models are necessarily confined to the invariant subspace {\bf HYP} in Table 1, defined by the intersection of (\ref{shxh2a})--(\ref{shxh2c}) and $0<\Omega_q<1$. 

\subsection{Critical points.}

The critical points of the dynamical system (\ref{DS1})--(\ref{DS3}) ($\varepsilon =1$) are listed in Table \ref{tabla2} and displayed by Figure 2. These points are located in the invariant subspace {\bf HYP}. Notice that {\bf MIL} is a subset of the invariant subspace {\bf VAC} of Table 1 and {\bf BB} corresponds to a non--simultaneous Big Bang ($\tbb'\leq 0,\,\,\Rightarrow\,\,\Dd\leq 0$).

\subsection{Asymptotic behavior.}

All phase space trajectories evolve from the past attractor {\bf BB} towards the future attractor {\bf MIL} (see Table \ref{tabla2} and Figure 2), respectively corresponding to the early/late asymptotic limits $\xi\to-\infty$ and $\xi\to\infty$ (equivalently $a\to 0,\,t\to\tbb$ and $a\to \infty,\,t\to\infty$). Considering  (\ref{gmode})--(\ref{dmode}), (\ref{jg})--(\ref{jd}) and (\ref{Omxi}), the forms of $\Omega_q,\,\jg,\,\jd$ in these asymptotic limits are given by: 
\begin{itemize} 
\item Near {\bf BB}: ($\xi\ll -1,\,t\approx \tbb,\,a\ll 1$):
\bse\ba \fl \Omega_q \approx 1-\frac{1-\Omega_{q0}}{\Omega_{q0}}a \to 1,\label{Omhas1}\\
\fl \jg\approx -\frac{2(1-\Omega_{q0})\Dig}{5\Did\Omega_{q0}^{3/2}}a^{5/2}\to 0,\quad  \Jg \approx \frac{2(1-\Omega_{q0})\Dig}{5\Omega_{q0}}a\to 0,\label{jghas1}\\
\fl \jd \approx -1+\frac{1}{3\Did\HH_{q0}\sqrt{\Omega_{q0}}}a^{3/2}\to -1,\quad \Jd\approx \frac{3\Did\HH_{q0}\sqrt{\Omega_{q0}}}{a^{3/2}}\to -\infty,\label{jdhas1}\ea\ese
\item Near {\bf MIL}: ($\xi\gg 1,\,t\to\infty,\,a\to\infty$):
\bse\ba \fl \Omega_q \approx \frac{\Omega_{q0}}{(1-\Omega_{q0})a}\to 0,\label{Omhas2}\\
\fl \jg\approx \frac{\Dig}{1-\Dig}+O((\ln\,a)/a),\qquad \Jg \approx \Dig+O((\ln\,a)/a),\label{jghas2}\\
\fl \jd\approx \frac{3\Did\sqrt{1-\Omega_{q0}}}{(1-\Dig)\,a}\to 0,\qquad \Jd\approx \frac{3\Did\sqrt{1-\Omega_{q0}}}{a}\to 0,\label{jdhas2}\ea\ese
\end{itemize}
where we have assumed in (\ref{jdhas1}) that $\Did\leq 0$ holds in order to fulfill (\ref{shxh2c}). The following remarks are worth highlighting:
\begin{itemize}
\item The limiting values of (\ref{Omhas1})--(\ref{jdhas1}) and (\ref{Omhas2})--(\ref{jdhas2}) fully coincide with the phase space coordinates of, respectively, {\bf BB} and {\bf MIL} given in Table \ref{tabla2}.
\item The nonzero asymptotic late time limit of $\jg$ in (\ref{jghas2}) (line of sinks {\bf MIL} in Table \ref{tabla2}) depends the choice of initial conditions ({\it i.e.} $\Dig$).
\item  The asymptotic forms in (\ref{Omhas1})--(\ref{jdhas1}) and (\ref{Omhas2})--(\ref{jdhas2}) yield for whatever choice of initial conditions at $\xi=0$:
\begin{equation} \fl |\Jd|\gg |\Jg|\quad \hbox{early times},\qquad |\Jg|\gg |\Jd|\quad \hbox{late times},\label{modsrel}\end{equation}
which clearly illustrate the expected early times dominance of the decaying mode and late times dominance of the growing mode (see Figure 1a). Considering (\ref{gmode}), (\ref{amplitudes}), (\ref{DgDd}) and (\ref{jg}), we have the following sign condition for each layer:
\begin{equation} \hbox{sign of}\,\,(\Dig) = \hbox{sign of}\,\, (\Dg,\,\Jg,\,\jg)\,\,\hbox{for all}\,\,\,\xi,\label{signJgh}\end{equation}
which defines the sign of $\Dg$ and the common sign of $\Jg$ and $\jg$ as invariant subspaces of $\PP$ for all trajectories with a given sign of $\Dig$.
\item As depicted by Figure 1a, we have $\Jd\to -\infty$ at the past attractor {\bf BB} and $\Jd=0$ at the future attractor {\bf MIL}, with $\Jd$ remaining negative for all $\xi$ because of $\HH_q>0$ and $\Did\leq 0$ (in compliance with (\ref{shxh})). On the other hand, $\Jg$ goes from zero at {\bf BB} to the finite terminal value (\ref{jghas2}) at the line of sinks {\bf MIL}, with the sign of this terminal value determined by the sign of $\Dig$.
\end{itemize}

\subsection{Phase space evolution.}

Figure 2 displays typical phase space trajectories together with invariant subspaces and the critical points listed in Table 2. The curves were generated by the analytic expressions (\ref{jg})--(\ref{jd}) and (\ref{Omxi}) for initial conditions complying with the HL conditions (\ref{shxh}), hence they are confined to the invariant subspace {\bf HYP} bounded by the wireframe and solid surfaces (yellow and grey in the online version) defined by the HL constraints (\ref{shxh2a})--(\ref{shxh2c}) (see also Table 1). Following the asymptotic limits (\ref{Omhas1})--(\ref{jdhas1}) and (\ref{Omhas2})--(\ref{jdhas2}), the phase space coordinates are restricted for all trajectories to
\begin{equation}\fl -1\leq \jd\leq 0,\qquad 0\leq |\jg| \leq \frac{|\Dig|}{1-\Dig},\qquad 0<\Omega_q<1,\end{equation}
where we remark that $1-\Dig\geq 0$ must hold in compliance with (\ref{shxh}). The curves evolve from the past attractor {\bf BB} at $\Omega_q=1$ (red circle), towards decreasing values of $\Omega_q$, avoiding the saddles {\bf EdS}, {\bf S1}, {\bf S2} (black circles) and the {\bf FLRW} subspace (thick dotted vertical line). Each curve terminates in a point in the line of sinks {\bf MIL} (subset of {\bf VAC}) in the plane $\Omega_q=0$. The position of each sink as terminal point in {\bf MIL} for each dust layer is given by the asymptotic value of $\jg$ (proportional to that of $\Jg$), and is completely determined by initial conditions through the sign of $\Dig$ (see section 7.4). As shown in section 11.4, the cases $\Dig\geq 0$ and $\Dig\leq0$ respectively yield asymptotic void and clump density profiles at the future attractor {\bf MIL} (see section 11.4). While initial conditions can be selected so that $\Dig$ changes sign for different ranges of $r$, the examples in Figure 2 have been selected so that $\Dig\geq 0$ (Figure 2a) and $\Dig\leq 0$ (Figure 2b) hold for all $r$, with trajectories complying with (\ref{signJgh}) for all $\xi$. Hence, these trajectories evolve within invariant subspaces defined by the sign of $\Dig$. 
\section{The general case: elliptic models.}
\begin{figure}
\begin{center}
\includegraphics[scale=0.4]{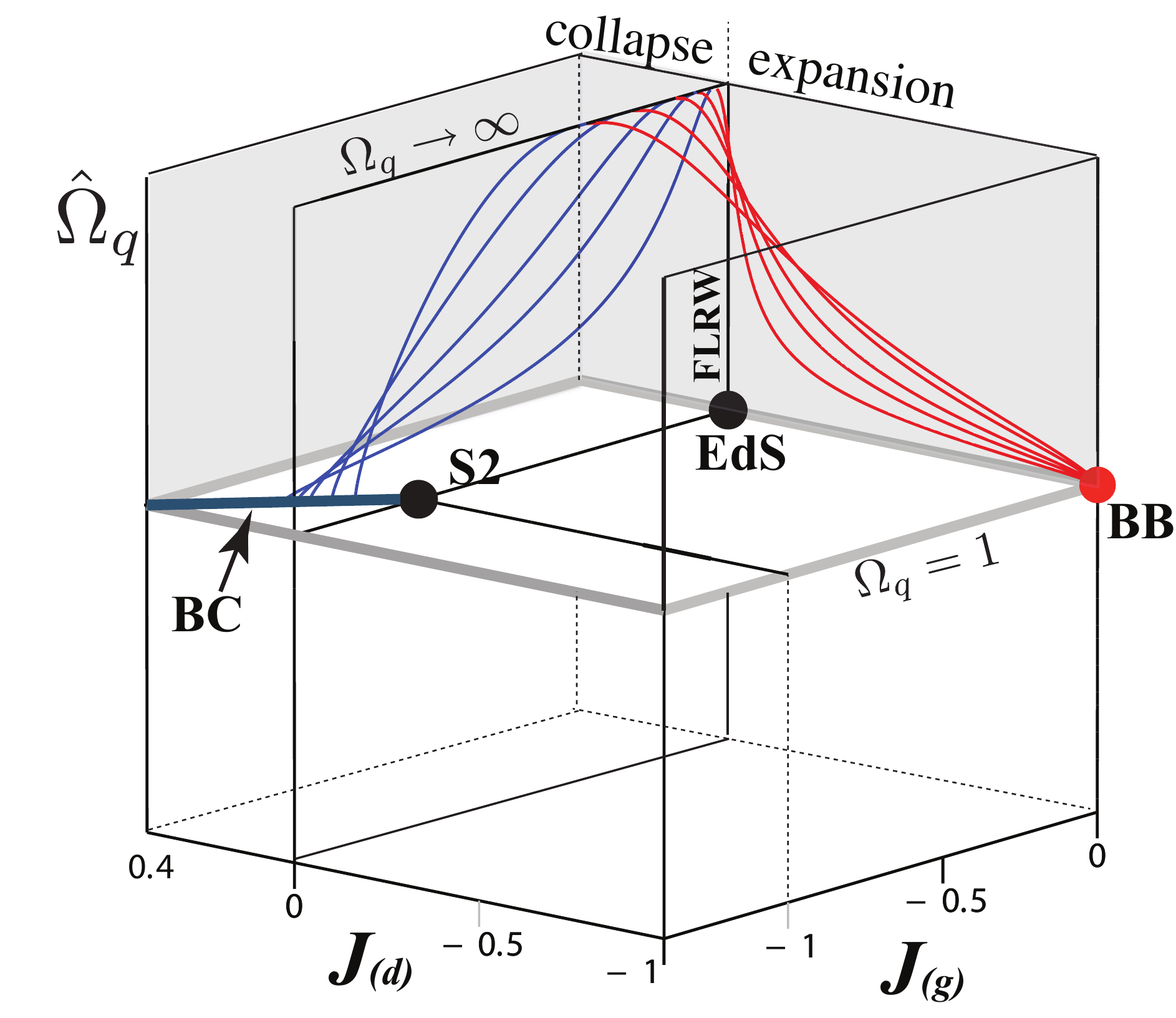}
\caption{{\bf Phase space evolution of the general case: elliptic models.} The figure displays phase space trajectories in the expanding and collapsing stages (in the online version expanding and collapsing curves are respectively depicted in red and blue). Critical points are listed in Table 2. The vertical axis variable is $\hat\Omega_q=\arctan \Omega_q$. The details of the evolution of phase space trajectories and description of critical points and invariant subspaces are discussed in section 8.4. All trajectories are confined in the invariant subspace {\bf ELL}.}
\label{fig1}
\end{center}
\end{figure}
\subsection{Absence of shell crossings.}

Phase space trajectories are contained in the invariant subspace $\Omega_q>1$, but, as with regular hyperbolic models, the HL conditions to avoid shell crossings necessarily impose further restrictions on the region of $\PP$ where the trajectories are confined. The HL conditions for elliptic models are given by the following constraints among initial conditions \cite{ltbstuff,RadAs,RadProfs}:
\begin{equation} \fl 1+\Drho_0\geq 0, \qquad \tbb'\leq 0\;\;(\Rightarrow\;\; \Did\leq 0),\qquad \tcoll'\geq 0.\label{shxe}\end{equation}
The first two conditions above yield the same constraints as the first and third constraints in (\ref{shxh}), while the joint fulfillment of the second and third condition above can be rephrased as the constraint:
\begin{equation}\fl \frac{r}{3}\tcoll' = \Tqcoll\Dbe_0 -\tau_{q0}\DT_0 = (1+\Drho_0)(2\pi\beta_{q0}\Dig-|\Did|)\geq 0.\label{shxe1}\end{equation}
where we used the form for $\tcoll$ in (\ref{rTqe2}) and $\Did\leq 0$ from (\ref{shxe}). Notice that (\ref{shxe1})  identifies $\Dig\geq 0$ as a necessary but not sufficient HL condition. Therefore, as a consequence of (\ref{beta}), (\ref{amplitudes}), (\ref{idents}) and (\ref{DgDd}), the HL conditions (\ref{shxe}) and (\ref{shxe1}) lead to the following constraints preserved by the fluid flow that must hold for the full evolution time range $\tbb<t<\tcoll$:
\bse\ba  \fl 1+\Drho\geq 0\quad \Rightarrow\quad \jg+\jd+1\geq 0,\label{shxe2a}\\
\fl 2\pi\beta_q\Dg+\Dd\geq 0 \quad \Rightarrow\quad \jd -\frac{\pi\Omega_q}{(\Omega_q-1)^{3/2}(\frac{2}{3}-W_q)}\jg\geq 0,\label{shxe2b}\\
\fl \Dg\geq 0\quad\Rightarrow\quad \jg\leq 0,\label{shxe2c}\\
\fl \Dd\leq 0 \quad\Rightarrow\quad \jd \left\{ \begin{array}{l}
 \leq 0,\qquad  
 {\hbox{expanding phase}}\quad \HH_q>0,\\ 
 \geq 0, \qquad 
 {\hbox{collapsing phase}}\quad \HH_q<0,\\ 
 \end{array} \right.,\label{shxe2d}
\ea\ese
where $W_q$ is defined by (\ref{Wqdef}) and we used the fact that $\HH_q\tau_q-2/3<0$ holds for the full time evolution, as well as using (\ref{rhoq})--(\ref{perts2}), (\ref{beta}), (\ref{DrhoJ}), (\ref{DK}), (\ref{jg})--(\ref{jd}) and (\ref{Dh2})--(\ref{DOm2}) to eliminate $\beta_q,\,\Drho$ and $\DKK$ in terms of $\Omega_q,\,\jg,\,\jd$. Notice that, as depicted by Figure 1b,  the HL conditions imply that both density modes are negative in the expanding stage, while $\Jg\leq 0$ and $\Jd\geq 0$ in the collapsing stage.  

As with hyperbolic models, each one of the HL constraints above is an invariant subspace of $\PP$, hence, the invariant subspace of $\PP$ associated with regular elliptic models (denoted by {\bf ELL} in Table 1) is the intersection of $\Omega_q>1$ and each one of (\ref{shxe2a})--(\ref{shxe2d}). All phase space trajectories of regular elliptic models are thus confined to {\bf ELL}. The critical points are listed in listed in Table \ref{tabla2}. Some of these points emerge in the dynamical system associated with the expanding stage and others from that of the collapsing stage.  

\subsection{Expanding stage.}

The dynamical system is the same as that of hyperbolic models, but with $\Omega_q>1$ and $W_q=Y_q$ given by (\ref{Y}) with $\epsilon=-1$ and $\ACal=$ arccos. The critical points (listed in Table \ref{tabla2}) are: the past attractor {\bf BB} and the saddles {\bf EdS} and {\bf S2}.  

\subsubsection{Asymptotic behavior.}
\begin{itemize}
\item {\underline{Near the past attractor}} {\bf BB}.  The phase space variables have the same limiting values (\ref{Omhas1})--(\ref{jdhas1}) of hyperbolic models.
\item {\underline{Near the maximal expansion.}} Late time evolution can be associated with the maximal expansion $\HH_q\to 0$ as $\xi\to \ln [\amax]$ (equivalently $t\to \tmax,\,a\to \amax=\Omega_{q0}/(\Omega_{q0}-1)$). Phase space variables in this limit take the form:
\bse\ba \fl \Omega_q=\frac{\amax}{(\Delta a)^2}\to \infty,\label{Ommax}\\
\fl \jg \approx \frac{-2\Dig}{1+2\Dig}+O(\Delta a),\qquad \Jg\approx -2\Dig+O(\Delta a),\label{jgmax}\\
\fl \jd \approx \frac{3\Did\HH_{q0}(\Omega_{q0}-1)^2\,\Delta a}{(1+2\Dig)\Omega_{q0}^{3/2}}\to 0,\qquad \Jd\approx \frac{3\Did\HH_{q0}(\Omega_{q0}-1)^2\,\Delta a}{\Omega_{q0}^{3/2}}\to 0, \nonumber\\ \fl \label{jdmax}\ea\ese
where $\Delta a\equiv \sqrt{\amax-a}\approx 0$ and $1+2\Dig\geq 0$ must hold because of the HL conditions (\ref{shxe}) and (\ref{shxe1}).
\end{itemize}

\subsection {Collapsing stage.} 

\subsubsection{Critical points.} 

The appropriate dynamical system is (\ref{DS1})--(\ref{DS3}) with $\Omega_q>1,\,\varepsilon=-1$ and $W_q$ given by (\ref{Wqdef}) with $\HH_q<0$. Its critical points (listed in Table \ref{tabla2}) are the following: the {\bf EdS} saddle and {\bf Big Crunch}, the latter defining a line of future attractors associated with the collapsing singularity reached by dust layers as $\xi\to\infty$ (or equivalently $a\to 0,\,\Omega_q\to 1$ as $t\to\tcoll$). 

\subsubsection{Behavior near the Big Crunch.}

It is evident, from its phase space coordinates in Table \ref{tabla2}, that the future attractor (the line of sinks {\bf BC}) is radically different from the past attractor (source) {\bf BB} of the expanding stage. This fact clearly indicates that the dynamics of dust layers near the collapsing singularity is qualitatively different from their behavior the initial big bang of the expanding stage.  

Considering the appropriate forms of $\jg,\,\jd$ in (\ref{jg}) and (\ref{jd}) and the density modes in (\ref{gmode}) and (\ref{dmode}), we have near the future attractor {\bf BC}: 
\bse\ba \fl \jg \approx \frac{-\pi \Dig \Omega_{q0}}{\Did(\Omega_{q0}-1)^{3/2}\HH_{q0}+\pi\Dig\Omega_{q0}}+O(\sqrt{a}),\quad \Jg\approx \frac{-3\pi\Dig\Omega_{q0}^{3/2}}{(\Omega_{q0}-1)^{3/2}\,a^{3/2}}\to-\infty,\nonumber\\
\label{BCg}\\
\fl \jd\approx \frac{-\Did\HH_{q0}(\Omega_{q0}-1)^{3/2}}{\Did(\Omega_{q0}-1)^{3/2}\HH_{q0}+\pi\Dig\Omega_{q0}}+O(\sqrt{a}),\qquad \Jd\approx \frac{-3\Did\HH_{q0}\sqrt{\Omega_{q0}}}{a^{3/2}}\to\infty,\nonumber\\
\label{BCd}\ea\ese
where we took under consideration that $\Dig\geq 0$ and $\Did\leq 0$ hold in compliance with (\ref{shxe2c}) and (\ref{shxe2d}). The following points are worth remarking:
\begin{itemize}
\item The finite nonzero asymptotic forms of $\jg$ and $\jd$ in (\ref{BCg})--(\ref{BCd}) satisfy the constraint $\jd+\jg=-1$ and provide the initial conditions dependent value of $\jg$ in Table \ref{tabla2}. Since $\jg+\jd=\Drho$, these asymptotic forms are consistent with the known behavior $\Drho\to -1$ as trajectories reach the Big Crunch {\bf BC} (see \cite{sussDS1,sussDS2}).    
\item The HL conditions (\ref{shxe2c})--(\ref{shxe2d}) imply that $\jg,\,\Jg\leq 0$ (as in the expanding stage), but $\jd,\,\Jd\geq 0$ (see Figure 1b). While both $\Jg$ and $\Jd$ diverge near {\bf BC}, the ratio of their magnitudes in (\ref{BCg})--(\ref{BCd}) together with the HL condition (\ref{shxe1}) imply that the growing mode is dominant over the decaying mode in this limit:
\begin{equation} \fl \left| \frac{\jd}{\jg}\right|=\left| \frac{\Jd}{\Jg}\right|= \frac{|\Did|}{\Tqcoll\,\Dig}=\frac{|\Dd|}{\Tqcoll\,\Dg}\leq 1\qquad \hbox{as}\quad \xi\to\infty \,\,(t\to\tcoll),\label{collratio}\end{equation}
which re--affirms a late time behavior $|\Jg|\geq |\Jd|$ near {\bf BC} that is qualitatively different from the early time behavior $|\Jd|\gg |\Jg|$ near the big bang in (\ref{modsrel}).
\end{itemize}

\subsection{Phase space evolution.}

The phase space trajectories of typical dust layers in the expanding and collapsing stage is depicted by Figure 3, where we use the variable $\hat\Omega_q\equiv \arctan\Omega_q$ (vertical axis) to associate the finite value $\hat\Omega_q=\pi/2$ to $\Omega_q\to\infty$. Expanding and collapsing trajectories are respectively depicted in red and blue in the online version and critical points are listed in Table 2. The curves were obtained from the analytic expressions (\ref{jg})--(\ref{jd}) and (\ref{Omxi}), for initial conditions complying with the HL conditions (\ref{shxe}), hence they are confined to the invariant subspace {\bf ELL} bounded by surfaces of $\PP$ defined by the HL constraints (\ref{shxe2a})--(\ref{shxe2d}) (these surfaces are not displayed by figure 3). The curves evolve expanding from the Big Bang past attractor {\bf BB} at $\Omega_q=1$ (red circle), which is common to hyperbolic models, towards increasing values of $\Omega_q$, avoiding the saddles {\bf EdS}, {\bf S2} (black circles) and the {\bf FLRW} subspace. They reach maximal expansion as $\Omega_q\to \infty,\,\hat\Omega_q=\pi/2$. In the collapsing stage they evolve towards decreasing $\Omega_q$, with each curve terminating in a point in the Big Crunch line of sinks {\bf BC} (see Table \ref{tabla2}) in the plane $\Omega_q=1$ (dark blue thick line). The position of each sink as terminal point in {\bf BC} for each dust layer is given by the asymptotic values of $\jg$ and $\jd$ given by (\ref{BCg}) and (\ref{BCd}), and is completely determined by initial conditions.
\begin{figure}
\begin{center}
\includegraphics[scale=0.4]{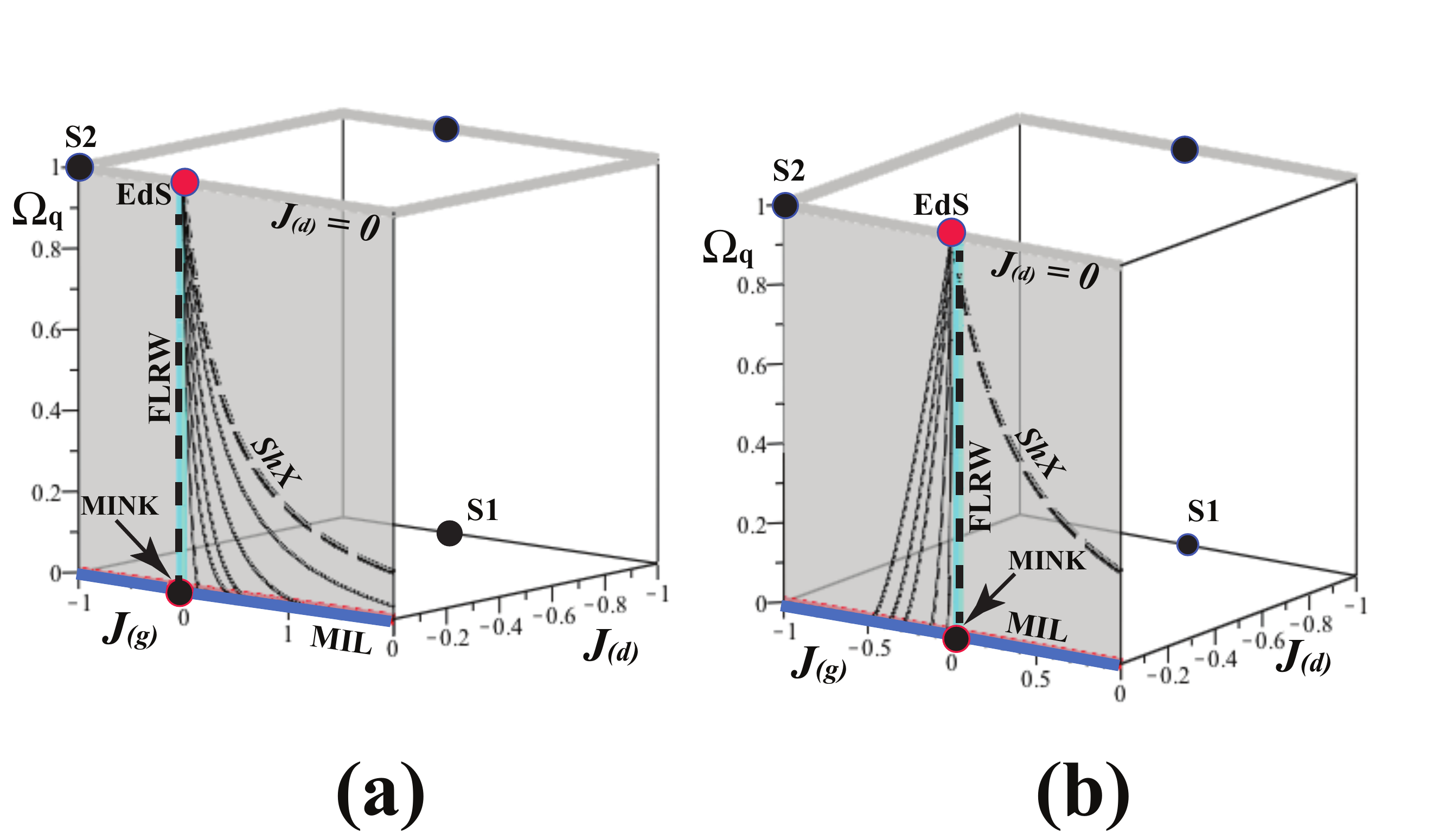}
\caption{{\bf Phase space evolution of regular hyperbolic models with suppressed decaying mode.} The panels (a) and (b) respectively depict the cases with $\Dig\geq 0$ and $\Dig\leq 0$ (so that $\jg\geq 0$ or $\jg\leq 0$ hold for all trajectories). Details of the phase space evolution, critical points are listed in Table 3 and invariant subspaces are discussed in section 9. All trajectories are confined to the invariant subspace {\bf SDM} (shaded plane $\jd=0$) and within the curve {\bf ShX} that marks the intersection of this invariant subspace and {\bf HYP} (the constraint (\ref{shxhSGM})). Notice that the Einstein de Sitter point {\bf EdS} is no longer a saddle, but the past attractor marking a homogeneous asymptotic early time regime.}
\label{fig1}
\end{center}
\end{figure}
\begin{figure}
\begin{center}
\includegraphics[scale=0.4]{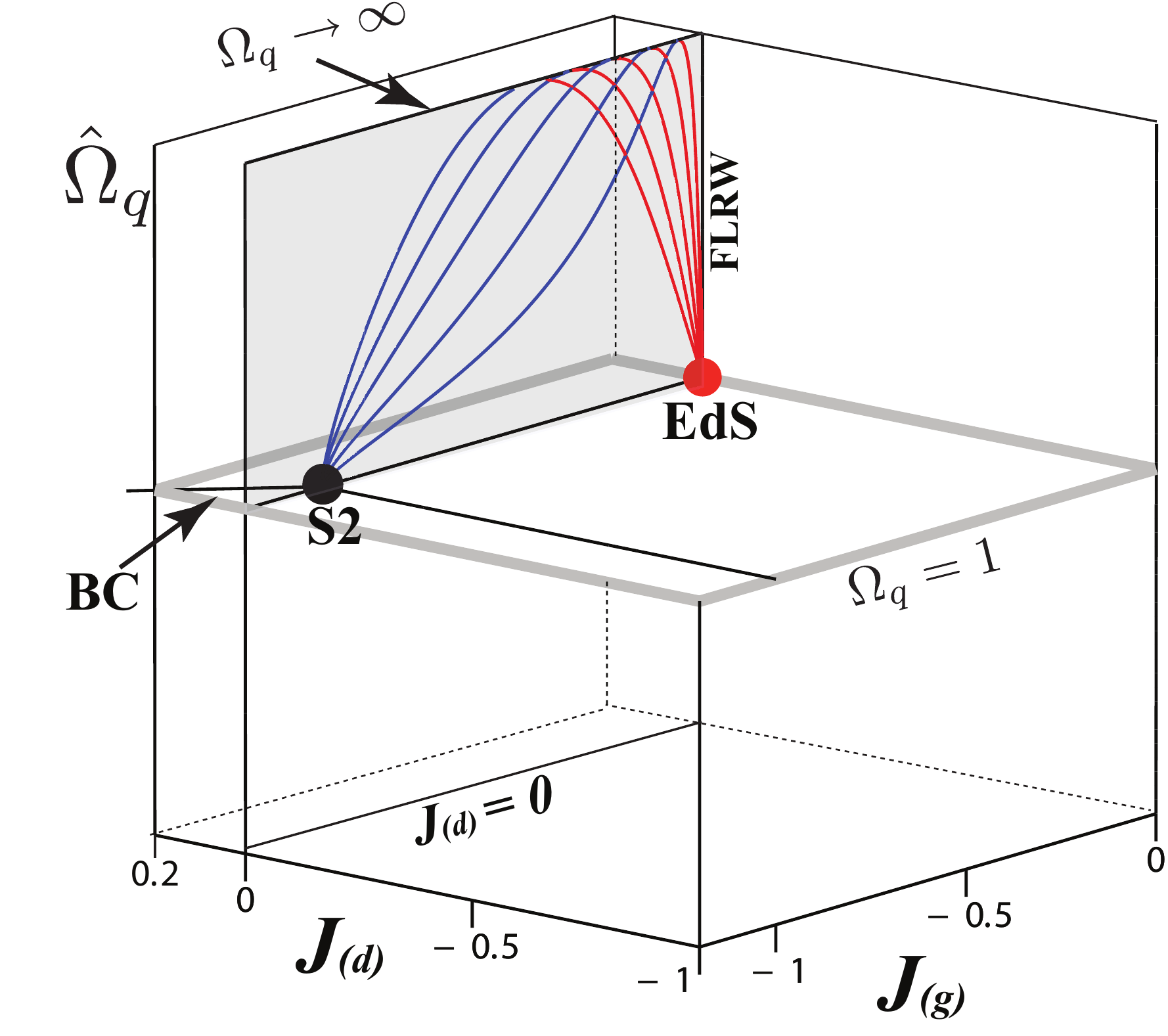}
\caption{{\bf Phase space evolution of regular elliptic models with suppressed decaying mode.} As in Figure 3, the vertical axis is parametrized by $\hat\Omega_q=\arctan\Omega_q$. The trajectories are confined to the intersection of the invariant subspace {\bf SDM} (shaded plane $\Jd=0$) and {\bf ELL}, hence $\jg\leq 0$ and $\Omega_q>1$. Details of the phase space evolution are discussed in section 9 and critical points are listed in Table 3. As with hyperbolic models, the Einstein de Sitter point {\bf EdS} is no longer a saddle, but the past attractor marking a homogeneous asymptotic early time regime.}
\label{fig1}
\end{center}
\end{figure}
\section{Switching off the decaying mode.}

%
\begin{table}
\begin{center}
\begin{tabular}{|c| c| c|}
\hline
\hline
\hline
\hline
\multicolumn{3}{|c|}{{\bf Critical Points. Suppressed Decaying Mode:} $\jg\ne 0,\,\jd = 0$.}
\\
\hline
\hline
\hline
\hline
\multicolumn{3}{|c|}{Hyperbolic models. Section 9. Figure 4.}
\\  
\hline
\hline
\hline
\hline
{Symbol} &{Phase Space coordinates} &{Description} 
\\
\hline
{{\bf EdS}} &{$\Omega_q=1,\,\jd=0,\,\jg=0$} &{Einstein--de Sitter. Past attractor}
\\  
{} &{} &{ (source). Simultaneous Big Bang}
\\
\hline  
{{\bf MIL}} &{$\Omega_q=0,\,\jd=0,\,\jg=\frac{\Dig}{1-\Dig}$} &{Milne. Future attractors (sinks)} 
\\ 
\hline
{{\bf S1}} &{$\Omega_q=0,\,\jd=-1,\,\jg=0$} &{Saddle}
\\ 
\hline
\hline
\hline
\hline
\multicolumn{3}{|c|}{Elliptic models. Section 9. Figure 5.}
\\
\hline
\hline
\hline
\hline
{Symbol} &{Phase Space coordinates} &{Description} 
\\
\hline
{{\bf EdS}} &{$\Omega_q=1,\,\jd=0,\,\jg=0$} &{Einstein--de Sitter. Past attractor}
\\  
{} &{} &{ (source). Simultaneous Big Bang}
\\    
\hline
{{\bf S2}} &{$\Omega_q=1,\,\jd=0,\,\jg=-1$} &{Future attractor (sink), ${\textrm{{\bf S2}}} \subset {\textrm{{\bf BC}}}$.}
\\    
\hline
\hline
\hline
\hline
\end{tabular}
\end{center}
\caption{{\bf{Critical points of $\PP$ for models with suppressed decaying mode}}. These critical points (displayed in Figures 4 and 5) follow from the dynamical system (\ref{DS1a})--(\ref{DS2a}), with $\varepsilon =1$ for hyperbolic models and elliptic models in their expanding phase ($\HH_q>0$) and $\varepsilon =-1$ for the collapsing stage ($\HH_q<0$) of the latter models.}
\label{tabla3}
\end{table}

Suppression of the decaying mode yields hyperbolic and elliptic models
\footnote{Models with a suppressed decaying mode cannot be parabolic, since $\tbb'=0$ is inly compatible with $\Drho_0=\Dh_0=0$ (FLRW limit) if $\Omega_{q0}-1=\KK_{q0}=0$ (see the following section).}
with initial conditions selected so that $\Did=0$, which is equivalent to demanding $\tbb'=0$ (or equivalently $\tau'_q=\DT=\DT_0=0$), which leads to a simultaneous big bang: $\tbb=\tbbo=$ constant. Initial conditions for these models follow from the analytic solutions (\ref{hypsol})--(\ref{ellsol}):
\begin{equation}\HH_{q0}=\frac{W_{q0}}{t_0-\tbbo},\qquad \Dh_0 = \frac{\Omega_{q0}}{W_{q0}}\frac{\dd W_{q0}}{\dd\Omega_{q0}}\DOm_0,\end{equation}
where $W_{q0}=W_q(\Omega_{q0})$ with $W_q$ given by (\ref{Wqdef}), so that the remaining initial value functions can be obtained from (\ref{rhoq})--(\ref{perts2}) evaluated at $t=t_0$. It is evident that these models can be fully determined by specifying a single free parameter, which can be $\Omega_{q0}$ in order to provide the easiest representation. 

Models with a suppressed decaying mode evolve in the invariant subspace {\bf SDM} in Table 1, which is the plane $\jd=0$ parametrized by $[\jg,\,\Omega_q]$ with $\Omega_q\ne 1$. However, the HL constrains (\ref{shxh2a})--(\ref{shxh2b}) and (\ref{shxe2a})--(\ref{shxe2c}) further restrict the region of phase space evolution: if $\jd=0$ these constraints reduce to:
\bse\ba \fl 1+\jg\geq 0,\qquad (1+\jg)Y_q-\left(\frac{2}{3}+Y_q\right)\geq 0,\qquad \hbox{hyperbolic},\label{shxhSGM}\\
\fl -1\leq \jg\leq 0,\qquad \hbox{elliptic},\label{shxeSGM}\ea\ese 
and define invariant subspaces in the plane $[\jg,\,\Omega_q]$ in which phase spaces trajectories are confined (we used the fact that $W_q=Y_q$ for hyperbolic models). 

The dynamical system (\ref{DS1})--(\ref{DS3}) also becomes reduced: the differential equation (\ref{DS3}) becomes a constraint that is identically satisfied, with the two remaining equations (\ref{DS1})--(\ref{DS2}) leading to the 2--dimensional dynamical system
\bse\ba \fl \frac{\partial\Omega_q}{\partial\xi} = \varepsilon \,\Omega_q(\Omega_q-1),\label{DS1a}\\
\fl\frac{\partial\jg}{\partial\xi} = \varepsilon\,\frac{[2-(\Omega_q+2)W_q]\jg(\jg+1)}{2(W_q-2/3)},\label{DS2a}\ea\ese
whose critical points are listed in Table \ref{tabla3}. 

The phase space evolution is depicted by Figures 4 and 5. The trajectories are confined to the invariant subspace {\bf SDM}. We have the following features: 
\begin{itemize}
\item {\underline {Hyperbolic models}} (Figure 4). The early time evolution is radically different from that of the general case, as trajectories emerge from the {\bf EdS} past attractor (red circle), but the late time evolution is qualitatively similar: curves evolve towards decreasing $\Omega_q$ and terminate at the future attractor (line of sinks) {\bf MIL} of the general case (dark blue thick line). The position of each curve in {\bf MIL} depends on initial conditions given by the terminal value (\ref{jghas2}). Curves with $\Dig\geq 0$ (Figure 4a) or $\Dig\leq 0$ (Figure 4b) respectively terminate with positive or negative values of $\jg$. As shown in section 11.4, initial conditions $\Dig\geq 0$ or $\Dig\leq 0$ respectively determine a void or clump density profile for the whole evolution.  
\item {\underline{Expanding stage of elliptic models}} (Figure 5). The curves are depicted in red in the online version. The early time evolution is qualitatively the same as that of hyperbolic models described above, but the curves evolve towards increasing $\Omega_q$ to a late time regime given by the maximal expansion $\hat\Omega_q=\arctan\Omega_q=\pi/2$ as $\xi\to \ln \amax$ analogous to the general case. 
\item {\underline {Collapsing stage of elliptic models.}} The curves are depicted in blue in the online version. They evolve from the maximal expansion $\hat\Omega_q=\pi/2$ towards the future attractor point {\bf S2}, contained in the line of sinks {\bf BC} of the general case. Again, the behavior near the collapse singularity is qualitatively different from that near the initial singularity in the expanding stage (past attractor {\bf EdS}).     
\end{itemize}
As we discuss in sections 11 and 12, the early time phase space behavior around the {\bf EdS} past attractor is the characterization of the early time homogeneity associated with models with suppressed decaying mode.
 
\section{Switching off the growing mode.}

%
\begin{table}
\begin{center}
\begin{tabular}{|c| c| c|}
\hline
\hline
\hline
\hline
\multicolumn{3}{|c|}{{\bf Critical Points. Suppressed Growing Mode:} $\jg= 0,\,\jd \ne 0$.}
\\
\hline
\hline
\hline
\hline
\multicolumn{3}{|c|}{Hyperbolic models: $\HH_q\tau_q>2/3$. Section 10. Figure 6.}
\\  
\hline
\hline
\hline
\hline
{Symbol} &{Phase Space coordinates} &{Description} 
\\
\hline
{{\bf BB}} &{$\Omega_q=1,\,\jd=-1,\,\jg=0$} &{Non--simultaneous Big Bang.}
\\  
{} &{} &{Past attractor (source)}
\\
\hline  
{{\bf MINK}} &{$\Omega_q=0,\,\jd=0,\,\jg=0$} &{Minkowski. Future attractor (sink)} 
\\
{} &{} &{${\textrm{{\bf MINK}}} \subset {\textrm{{\bf MIL}}}$ } 
\\
\hline  
{{\bf EdS}} &{$\Omega_q=1,\,\jd=0,\,\jg=0$} &{Einstein--de Sitter. Saddle}
\\  
\hline
{{\bf S1}} &{$\Omega_q=0,\,\jd=-1,\,\jg=0$} &{Saddle}
\\ 
\hline
\hline
\hline
\hline
\multicolumn{3}{|c|}{Parabolic models: $\HH_q\tau_q=2/3$.  Section 10. Figure 6.}
\\
\hline
\hline
\hline
\hline
{Symbol} &{Phase Space coordinates} &{Description} 
\\
\hline
{{\bf BB}} &{$\Omega_q=1,\,\jd=-1,\,\jg=0$} &{Non--simultaneous Big Bang.}
\\  
{} &{} &{Past attractor (source)}
\\
\hline
{{\bf EdS}} &{$\Omega_q=1,\,\jd=0,\,\jg=0$} &{Einstein--de Sitter. Future attractor (sink).}
\\    
\hline
\hline
\hline
\hline
\end{tabular}
\end{center}
\caption{{\bf{Critical points of $\PP$ for models with suppressed growing mode}}. The critical points of the hyperbolic models follow from the dynamical system (\ref{DS1aa})--(\ref{DS2ab}), while those for the parabolic models follow from the one--dimensional system (\ref{DSpar}). These points are displayed by Figure 6.}
\label{tabla4}
\end{table}

Models characterized by a suppressed growing mode $\jg=\Jg=0$ follow from either one of the following combination of restrictions:
\bse\ba \fl \Dig=0\,\,\hbox{with}\,\,\HH_q\tau_q\ne \frac{2}{3}\;\;\Rightarrow\;\; \beta_{q0}=\hbox{const.,}\;\;\hbox{(hyperbolic and elliptic models)},\label{zerog1}\\
\fl \Dig\ne 0\,\, \hbox{with}\,\,\HH_q\tau_q= \frac{2}{3}\;\;\Rightarrow\;\; \Omega_{q0}-1=\KK_{q0}=0\;\; \hbox{(parabolic models)}, \label{zerog2}\ea\ese
where $\beta_{q0}$ is defined in (\ref{beta}) and we used $\Dbe_0=(1+\Drho_0)\Dig$. These restrictions lead to the following set of restricted initial conditions:
\bse\ba \fl \beta_q = b_0=\hbox{const.}\;\; \Rightarrow\;\; \frac{4\pi}{3}\rho_q = b_0 |\KK_q|^{3/2} \;\;\hbox{or}\;\;  \HH_q = \frac{b_0\Omega_q}{2|1-\Omega_q|^{3/2}},\label{zerogIC1}\\ 
\fl \Omega_{q0}-1=\KK_{q0}=0 \;\; \Rightarrow\;\; \HH_{q}^2 =\frac{8\pi}{3}\rho_{q},\quad 2\Dh=2\Drho,\label{zerogIC2}\ea\ese
so that a single free parameter is sufficient to fully determine these models. 

The phase space evolution is contained in the invariant subspace {\bf SGM} in Table 1, which is the plane $\jg=0$ parametrized by $[\jd,\,\Omega_q]$, with the line $\Omega_q= 1,\,\jd=\jd(\xi)$ corresponding to parabolic models (the invariant subspace {\bf PAR} $\subset$ {\bf SGM} in Table 1). The HL constraints (\ref{shxe2a})--(\ref{shxe2d}) for elliptic models with $\Dig=0$ cannot be fulfilled, hence all regular models with a suppressed growing mode must be either hyperbolic models with $\Dig=\Dg=0$ or parabolic models, both complying with $-1\leq \jd\leq 0$ (from the HL constraints (\ref{shxh2a})--(\ref{shxh2c})).  

For hyperbolic models complying with $\Dig=\Dg=0$ the dynamical system (\ref{DS1})--(\ref{DS3}) reduces to the 2--dimensional system
\bse\ba \fl \frac{\partial\Omega_q}{\partial\xi} = \Omega_q(\Omega_q-1),\label{DS1aa}\\
\fl\frac{\partial\jd}{\partial\xi} = -\left(1+\frac{\Omega_q}{2}\right)\,\jd(\jd+1),\label{DS2ab}\ea\ese
whose critical points are listed in Table \ref{tabla4}.
   
For parabolic models ($\Omega_q=1$) the system (\ref{DS1aa})--(\ref{DS2ab}) further reduces to a single dynamical equation that determines the line {\bf PAR} in Table 1:
\begin{equation} \fl \frac{\partial\jd}{\partial\xi} = -\frac{3}{2}\jd(\jd+1)\quad\Rightarrow\quad \jd(\xi)=\frac{\jd{}_0}{(1+\jd{}_0)\textrm{e}^{3\xi/2}-\jd{}_0},\label{DSpar}\end{equation}
where $\jd{}_0=\jd(0,r)$ is given by (\ref{initconds2}). The two critical points of (\ref{DSpar}) are listed in Table \ref{tabla4}. 

The phase space evolution of models with a suppressed growing mode is depicted by figure 6. A qualitative description is provided below:
\begin{itemize} 
\item {\underline{Hyperbolic models.}} Because of the HL constraints, the phase space trajectories are contained in the restriction $-1\leq \jd\leq 0$ of {\bf SGW} (see Table 1), and thus, they are qualitatively analogous to those of the general case for early times near the past attractor {\bf BB}. However, the late time evolution is qualitatively different: all phase space trajectories terminate at the Minkowski sink {\bf MINK} associated with an asymptotic homogeneous vacuum state $\Dg,\,\Drho\to 0$ as $\xi\to\infty$.
\item {\underline{Parabolic models.}} They evolve in the invariant subspace {\bf PAR} from the Big Bang past attractor {\bf BB} ( $\xi\to-\infty,\,\jd\to -1$) to the future attractor {\bf EdS} ( $\xi\to\infty,\,\jd\to 0$). This future attractor singles out these models as the only LTB models whose asymptotic late time evolution is qualitatively analogous to that of an Einstein de Sitter model.
\end{itemize}
\begin{figure}
\begin{center}
\includegraphics[scale=0.4]{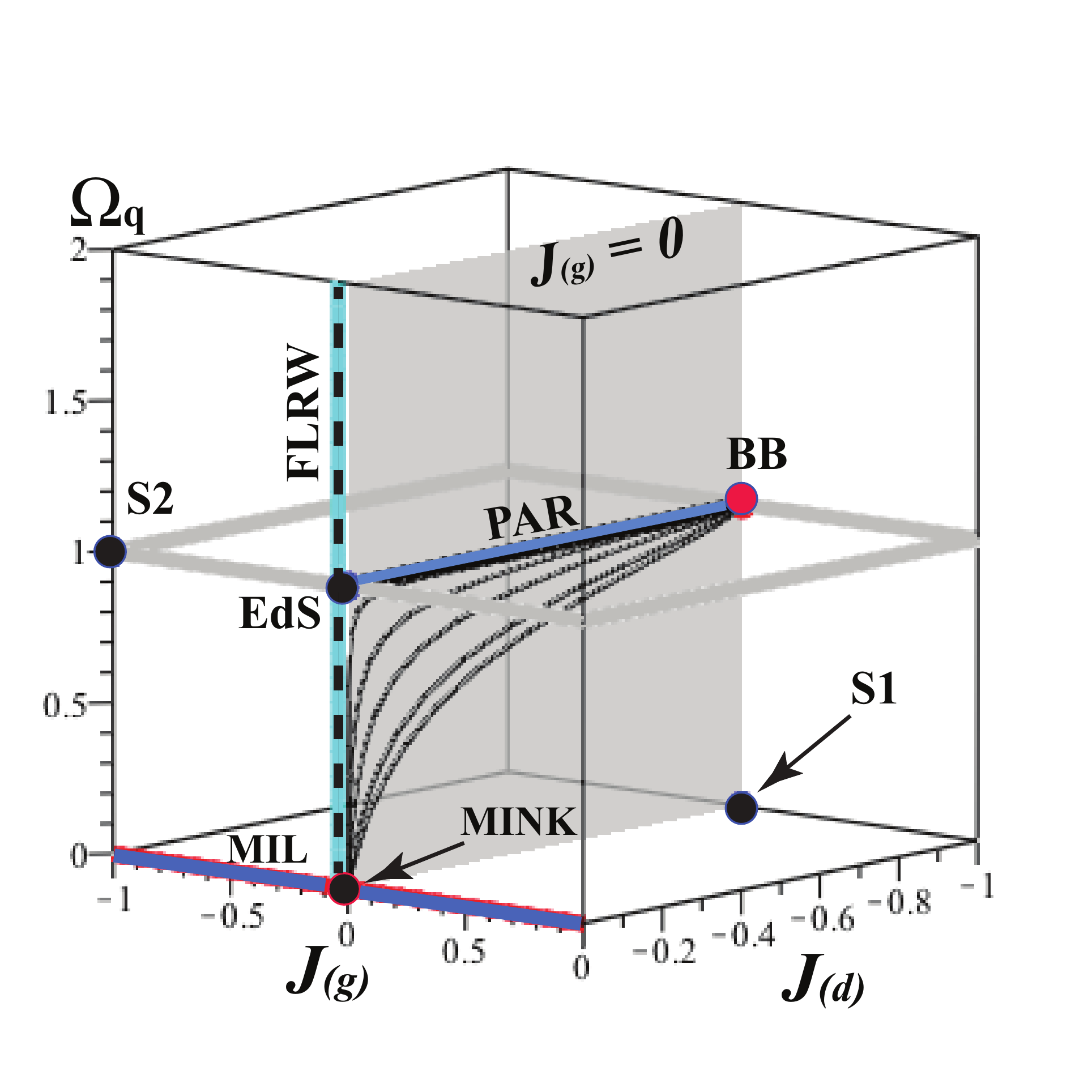}
\caption{{\bf Phase space evolution of regular models with suppressed growing mode.} The figure displays trajectories of hyperbolic models  contained in the region $0<\Omega_q<1$ of the invariant subspace {\bf SGM} (shaded plane $\jg=0$).  Elliptic models present shell crossings and their trajectories are not displayed. Parabolic models are confined to the invariant subset {\bf PAR} (the line $\Omega_q=1$) contained in {\bf SGM}. Details of the phase space evolution, critical points are listed in Table 4 and invariant subspaces are discussed in section 10. Notice that the Minkowski {\bf MINK} and Einstein de Sitter {\bf EdS} points are the future attractors of hyperbolic and parabolic models.}
\label{fig1}
\end{center}
\end{figure}

\section{Phase space evolution of the perturbations.}

Besides their coordinate independent nature \cite{part1}, the $\Da$ define a self--consistent gauge invariant perturbation formalism on a FLRW background associated with the q--scalars $\rho_q,\,\HH_q,\,\KK_q,\,\Omega_q$ (see \cite{part2} for a comprehensive discussion). Therefore, it is useful to examine the relation between the density modes and the phase space evolution of these perturbations (specially the density perturbation $\Drho=\jg+\jd$). 

\subsection{Early times regime.}   

If the decaying mode is nonzero ($\Did\leq 0$), this mode completely dominates the early time behavior of the perturbations, since $\jg,\,\Jg\approx 0$ and $\Drho\approx \jd$ hold near the past attractor {\bf BB} associated with a non--simultaneous big bang (these remarks also hold when the growing mode is suppressed $\jg=\Jg=0$). For whatever choice of initial conditions (\ref{initconds1})--(\ref{initconds2}), we have from  (\ref{DrhoJ}) and (\ref{DK})--(\ref{DOM}):
\bse\ba  \Drho \approx   -1- \frac{1}{3\sqrt{\Omega_{q0}}\,\Did}\,a^{3/2}\to -1,\label{Dma0}\\
\DKK \approx  -\frac{2}{3}\mp \frac{2[1+(3/2)\DKK_0]}{9\sqrt{\Omega_{q0}}\,\Did}\,a^{3/2}\to -\frac{2}{3},\label{Dka0}\\ 
 \Dh \approx  -\frac{1}{2} - \frac{\Omega_{q0}-1}{6\,\Omega_{q0}}\,a\to -\frac{1}{2},\label{DHa0}\\
 \DOm \approx -\frac{1-\Omega_{q0}}{3}\,a \to 0, \label{DOma0}
\ea\ese
which clearly illustrate how the dominance of the decaying mode is associated with a very inhomogeneous early times behavior of the scalars $A$ and $A_q$ since $|\Da|\ll 1$ does not hold (the exception being $\DOm$). This early time inhomogeneity can be appreciated by looking at the consequences of the limit $\Drho\to-1$ in (\ref{Dma0}): it indicates a very different behavior of the local density, $\rho$, and its associated q--scalar, $\rho_q$, in the early time regime:
\begin{equation}\rho \approx \frac{\rho_0}{3\sqrt{\Omega_{q0}}(-\Did)\,a^{3/2}}\ll \rho_q=\frac{\rho_{q0}}{a^3},\end{equation}
where we assumed that $\Did\leq 0$ holds (in compliance with (\ref{shxh})) and used $\rho=\rho_q(1+\Drho)$, as well as (\ref{DrhoJ}) and (\ref{Dma0}). While both densities diverge near {\bf BB}, they diverge at very different rates with $\rho\ll\rho_q$, so that $\rho/\rho_q=1+\Drho\to 0$, which provides a measure of inhomogeneity because $\rho_q$ can be associated with the density of an abstract FLRW background (see \cite{part2}). 

If the decaying mode is suppressed ($\Did=0$), the past attractor is no longer {\bf BB} but the {\bf EdS} point associated with a simultaneous big bang (see figures 4 and 5).  The early time behavior of the $\Da$ in the limit $\xi\to-\infty$ (or $a\to 0,\,t\to \tbbo$) is radically different from the corresponding limits when $\Did\leq 0$ in (\ref{Dma0})--(\ref{DHa0}): 
\ba  \fl \Drho \approx \frac{\epsilon\Dig\,\gamma_{q0}}{5}\,a\to 0,
 \qquad  \Dh\approx -\frac{\epsilon\,\Dig\,\gamma_{q0}}{15}\,a\to 0,\qquad
\DKK \approx \frac{-2\Dig}{3},\nonumber\\\label{Dsimbb}\ea
where $\epsilon=1,-1$ for hyperbolic and elliptic models, $\gamma_{q0}$ is defined in (\ref{Dal0}) and $\DOm$ having the same limit as $\Drho$ ({\it i.e.} (\ref{DOma0})). Evidently, the elimination of the decaying mode suppresses the early time inhomogeneity in the density and Hubble factor, as they decay at the same FLRW rate as their associated q--scalars: $\rho\approx \rho_q\sim a^{-3}$ and  $\HH\approx \HH_q\sim a^{-3/2}$, though this does not occur for the spatial curvature (unless we demand the extra condition $|\Dig|\ll 1$ in (\ref{Dsimbb})).    

Because of their compatibility with the near homogeneity expected for early Universe conditions, models with suppressed decaying mode have been preferred in most work dealing with observational applications (specifically void models, see \cite{BKHC2009,marranot} and references quoted therein).

\subsection{Late times regime.} 

If the growing mode is nonzero ($\Dig\ne 0$ and $\HH_q\tau_q\ne 2/3$), this mode  completely dominates the late time behavior of the perturbations, since $\jd,\Jd\approx 0$ and $\Drho\approx \jg$ hold, either near the future attractor {\bf MIL} of hyperbolic models ($t\to\infty,\,\xi\to\infty,\,\Omega_q\to 0$), or the maximal expansion ($t\to\tmax,\,\xi\to\ln\amax,\,\Omega_q\to\infty$) in elliptic models (see figure 1). The behavior of the density perturbation in both cases follows directly from (\ref{DrhoJ}):
\bse\ba \fl \Drho \approx \frac{\Dig}{1-\Dig},\qquad t\to\infty\;\;\hbox{(hyperbolic models)},\label{Drhoas1}\\
\fl \Drho \approx -\frac{2\Dig}{1+2\Dig},\qquad t\approx \tmax\;\;\hbox{(elliptic models)},\label{Drhoas2} \ea\ese 
with the perturbation $\DOm$ having the same late time limit as $\Drho$ in (\ref{Drhoas1}) for hyperbolic models (from (\ref{DOM})), while for elliptic models we have $\Dh\to\infty$ and $\DOm\approx -(\Omega_q-1)\HH_q\approx -\KK_q/\HH_q\to -\infty$ as $t\to\tmax$ (an expected result since $\Omega_q\to\infty$ and $\HH_q=0$ hold in this limit). The remaining perturbations $\DKK$ and $\Dh$ vanish near {\bf MIL}. Notice how the dominance of the growing mode yields an asymptotic nonzero value for $\Drho$ and $\DOm$ that is proportional to $\Jg\approx \Dig$, which indicates the existence of an inhomogeneous late time density (the ``residual inhomogeneity'' reported in \cite{wainwright}). 

\subsection{The collapsing regime.}

Near the future attractor {\bf BC} (see Table \ref{tabla2}), associated with the collapsing singularity (Big Crunch), the perturbations take the same asymptotic inhomogeneous forms (\ref{Dma0})--(\ref{DOma0}). However, the interpretation of these limits is radically different, as they are not associated with a dominant decaying mode, but a dominant growing mode. This follows from the fact that both modes $\Jg$ and $\Jd$ diverge in (\ref{BCg})--(\ref{BCd}) (see also Figure 1), with $|\Jg|>\Jd$ (from (\ref{collratio})), while $\jg$ and $\jd$ reach nonzero finite terminal values such that $\jg+\jd=\Drho=-1$. Another important difference is the fact that the limits (\ref{Dma0})--(\ref{DOma0}) also hold for elliptic models with a suppressed decaying mode as trajectories approach the future attractor {\bf S2} (see Figure 5 and section 9), while near the Big Bang past these limits only hold when the decaying mode is nonzero (the past attractor {\bf BB}). In a sense, the description of conditions near {\bf BC} allows us to disentangle the behavior near {\bf BC} from that near {\bf BB}, given the fact that the perturbations reach the same limits near both critical points. Hence, regardless of the existence of a decaying mode, there is a clear inhomogeneous growing mode dominated behavior near the Big Crunch that is qualitatively analogous to that of the Big Bang with a nonzero dominating decaying mode. 

\subsection{Density radial profiles.} 

As shown in \cite{RadProfs}, the relation between the gradients $\rho'_q,\,\rho'$ and the density perturbation $\Drho$ (see (\ref{Dadef})) leads to a close connection between the sign of $\Drho$ and the type of density radial profile: over density or ``clump'' if $\rho'_q\leq 0$ and under density or ``void'' if $\rho'_q\geq 0$ (for a comprehensive discussion see \cite{RadProfs}). The close relation between the density radial profiles and the growing mode emerges from the fact (see lemmas 7 and 9 and Table 1 of \cite{RadProfs}) that initial conditions that govern the evolution of these profiles depend on the sign of $\Dbe_0$ in (\ref{Dbe0}), which determines the sign of $\Dig,\,\Dg,\,\Jg$ and $\jg$ (see (\ref{gmode}), (\ref{amplitudes}), (\ref{jg}) and (\ref{DgDd})). Therefore, considering the early and late time limits of $\rho$ in (\ref{Dma0}), (\ref{Dsimbb}) and (\ref{Drhoas1})--(\ref{Drhoas2}), we arrive to the following general results relating density profiles and the sign of the density growing mode: 
\begin{itemize}
\item {\bf{If the growing mode is positive ($\Jg,\,\jg\geq 0$) a density void profile necessarily emerges in the late time evolution.}} This applies only to those  regular hyperbolic models with initial conditions $\Dig\geq 0$, irrespective of whether the decaying mode is zero or not (Figures 2a and 4a). 
\item {\bf{If the growing mode is positive ($\Jg,\,\jg\geq 0$) and the decaying mode is zero ($\Jg,\,\jg= 0$) a density void profile occurs for the whole time evolution.}} This applies only to those  regular hyperbolic models with initial conditions $\Dig\geq 0,\,\Did=0$ (Figure 4a).
\item  {\bf{If the growing mode is negative ($\Jg,\,\jg\leq 0$) a density clump profile necessarily occurs for the whole time evolution.}} This applies to all regular elliptic models (Figures 3 and 5) and to regular hyperbolic models with initial conditions $\Dig\leq 0$, irrespectively of whether the decaying mode is zero or not (Figures 2b and 4b). 
\end{itemize}
These general results involving the sign of the growing mode may seem to be counter--intuitive, because they are opposite to the ``conventional wisdom'' from the astrophysical literature \cite{kras2,BKHC2009,contrast},  whereby a positive growing mode is associated with an increasing ``density contrast'' in the formation of an over--density (clump), while a negative growing mode becomes intuitively connected to a decreasing ``density contrast'' in the formation of an under--density or void (see especially examples in \cite{kras2,BKHC2009,KH}, see \cite{part2} for a comprehensive discussion on this issue). 

The definition of ``clump'' or ``void'' profiles can be extended to other scalars \cite{part2,RadProfs}. For regular hyperbolic models it is straightforward to show that the pattern of profile evolution of $\Omega_q$ is qualitatively analogous to the profile evolution of $\rho$ and $\rho_q$, while as shown in Table 1 of \cite{RadProfs}, the profile pattern of $\KK_q$ is approximately the opposite to that of $\rho$ and $\rho_q$. The evolution of radial profiles of $\HH$ and $\HH_q$ exhibits a much weaker relation to the signs of the density models (see \cite{RadProfs}).    

\section{Inhomogeneity in terms of invariant scalars.}

The perturbations $\Da$ are covariant objects directly related to curvature and kinematic invariants \cite{part1}:
\begin{equation} \Drho = \frac{\phi}{1-\phi},\qquad \Dh = -\frac{\zeta}{1+\zeta},\label{invsc1}\end{equation}
\bse\ba 
 \fl  \phi \equiv \frac{6\Psi_2}{\RRR}=\frac{\Drho}{1+\Drho}=\Jg+\Jd,\label{invsc2}\\
 \fl \zeta \equiv \frac{\Sigma}{\HH}=-\frac{\Dh}{1+\Dh}=\frac{2(1-\Omega_q)\Dg-(\Omega_q+2)(\Jg+\Jd)}{6-2(1-\Omega_q)\Dg+(\Omega_q-4)(\Jg+\Jd)},\label{invsc3}\ea\ese
where $\RRR=8\pi\rho$ is the Ricci scalar, the scalars $\Psi_2,\,\Sigma$ are defined in (\ref{Sigma1}) and (\ref{EE1}) and we used the fact that $\Dg=\Dig$. Evidently, $\Drho$ and $\Dh$ provide an invariant measure of the deviation from FLRW homogeneity through the ratio of Weyl to Ricci curvature ($\phi$) and anisotropic to isotropic expansion ($\zeta$). The remaining perturbations $\DKK$ and $\DOm$ can also be expressed in terms of these invariants by substituting (\ref{invsc2}) and (\ref{invsc3}) into (\ref{perts1})--(\ref{perts2}) or (\ref{DK})--(\ref{DOM}). We have the following possibilities on the relation between $\phi,\,\zeta$ and the density modes:
\begin{itemize}
\item If the decaying mode is not suppressed ($\Did\leq 0$), these ratios take the following early time forms near the past attractor {\bf BB} 
\begin{equation}\fl \phi \approx \Jd\approx \frac{3\sqrt{\Omega_{q0}}\,\Did}{a^{3/2}}\to-\infty,\qquad \zeta \approx 1 -\frac{2(\Omega_{q0}-1)}{\Omega_{q0}}\,a\to 1, \label{ratio1}\end{equation}
where we assumed that $\Did\leq 0$ holds in compliance with (\ref{shxe}).
\item If the decaying mode is suppressed ($\Did=0$), then both ratios vanish near the past attractor {\bf EdS}. 
\item If the growing mode is not suppressed, the ratio $\phi$ takes the following forms in late time regime:  
\bse\ba \phi \approx \Dig\approx \Jg\quad \hbox{(near future attractor {\bf MIL})},\label{ratio2a}\\ \phi \approx -2\Dig\approx \Jg\quad\hbox{(near maximal expansion)},
\label{ratio2b} \ea\ese 
with the ratio $\zeta$ vanishing near {\bf MIL} and diverging near maximal expansion (because $\HH_q=0$ and $\Omega_q\to\infty$). 
\item If the growing mode is suppressed ($\Dig=0$), then both ratios take the same form as (\ref{ratio1}) near {\bf BB} and both vanish in the future attractor {\bf MINK}. 
\item In the collapsing regime $\zeta$ has the same form as in (\ref{ratio1}) near {\bf BC}, but $\phi$ takes the form:
\begin{equation}\fl \phi \approx  \frac{3\sqrt{\Omega_{q0}}\,[\pi\Dig\Omega_{q0}+\Did\HH_{q0}(\Omega_{q0}-1)^{3/2}]}{(\Omega_{q0}-1)^{3/2}\,a^{3/2}}\to\infty,\label{ratio3}\end{equation}
where we used (\ref{shxe1}) and the form for $\tcoll$ in (\ref{rTqe2}). 
\end{itemize}
It is important to remark that in most cases above $\phi$ increases as the evolution proceeds and the growing mode becomes dominant, the exceptions to this rule are the following two cases with suppressed decaying mode: hyperbolic models with $\Dig\leq 0$ and elliptic models, since $\phi$ goes from $\phi=0$ at the past attractor {\bf EdS} towards negative late time values (\ref{ratio2a})--(\ref{ratio2b}) (notice that $\Dig\geq 0$ is a necessary HL condition for elliptic models).         

\section{Summary and final discussion.}

We have found analytic exact covariant expressions ($\Jg,\,\Jd$ in (\ref{gmode}) and (\ref{dmode}), section 4) that generalize for LTB dust models the density growing and decaying modes of linear perturbation theory of dust sources (section 5). To achieve this task we considered the exact density perturbation, $\Drho$, that emerges from the description of LTB dynamics furnished by the quasi--local (q--scalars) $A_q$ and their local perturbations $\Da$, which were studied comprehensively in \cite{part1,part2} (the necessary background material appears in sections 2 and 3). As shown in these references, the q--scalars and their perturbations are themselves covariant scalars related to curvature and kinematic invariants (see section 12), and in particular the $\Da$ yield a rigorous self--consistent formalism of exact perturbations in which a ``FLRW background'' is defined by the $A_q$. Since this formalism reproduces in the linear limit the results of linear perturbation theory (in the comoving gauge), we show in section 5 that the linear limit of the exact growing and decaying modes are consistent with their corresponding forms obtained in previous literature \cite{zibin} for linear perturbations on an Einstein--de Sitter background.   

The relation between $\Jg,\,\Jd$  and the dynamical behavior of LTB models was studied thoroughly by means of a dynamical system defined in section 6, whose associated 3--dimensional phase space $\PP$ is parametrized by the q--scalar $\Omega_q$ (which generalizes the FLRW Omega factor) and two variables $\jg,\,\jd$ (see (\ref{jg})--(\ref{jd})) that are closely related to $\Jg,\,\Jd$. Hence, dust layers are curves (phase space trajectories)  evolving between the critical points in $\PP$ (listed in Tables 2, 3 and 4), with the full set of trajectories of each model defining a unique 2--dimensional surface in $\PP$. This phase space study, which is applicable to any LTB model admitting at least one symmetry center, represents an important improvement over previous work using a dynamical systems approach to LTB models \cite{sussDS1,sussDS2}, since: (i) we now describe the Hellaby--Lake (HL) regularity conditions (absence of shell crossing singularities) as fluid preserved constraints that are effectively invariant subspaces of $\PP$, and (ii) we provide a full description of the expanding and collapsing regimes of elliptic models in a unified phase space.   The phase space approach developed in this paper also represents a significant improvement over previous articles that used other means to obtain exact expressions for the density modes \cite{kras2,silk,wainwright} (see Appendix A for a critical review of this literature). 

Considering the HL conditions as fluid preserved constraints, the density modes provide an effective classification of regular LTB models in 4 non--vacuum subclasses that define regions of $\PP$ that are invariant subspaces, so that all phase space trajectories of any of these subclasses are entirely confined to its corresponding region. These 4 subclasses were examined separately: hyperbolic and elliptic models in the general case when both density modes are nonzero (subspaces {\bf HYP} and  {\bf ELL} discussed in sections 7 and 8), and models in which one of the modes is suppressed: suppressed decaying mode (invariant subspace {\bf SDM}, section 9) and suppressed growing mode (invariant subspace {\bf SGM}, which contains parabolic models {\bf PAR}, section 10). The main results of this phase space study are: 
\begin{itemize}
\item For all LTB models of the general case the early time evolution is governed by the decaying mode ($|\Jd|\gg|\Jg|,\,\,|\jd|\gg|\jg|$), whereas in the late time evolution (even in the collapsing stage) the growing mode is dominant ($|\Jg|\gg|\Jd|,\,\,|\jg|\gg|\jd|$). The signs of the modes are determined by the fulfillment of the HL conditions. The time evolution of $\Jg$ and $\Jd$ for a typical dust layer was displayed in Figure 1.
\item For all models in which the decaying mode is nonzero ($\jd\leq 0$ at early times from the HL conditions) the phase space trajectories begin their evolution in a past attractor {\bf BB} (see sections 7, 8 and 10, as well as Figures 2, 3 and 6 and Tables 2 and 4), associated with a non--simultaneous Big Bang and very inhomogeneous early time conditions (see discussion in sections 11.1 and 12).
\item If the decaying mode is suppressed, the past attractor is the Einstein de Sitter point {\bf EdS}, associated with a simultaneous Big Bang and early time homogeneous conditions (see sections 9, 11.1, Figures 4 and 5 and Table 3).
\item Phase space trajectories of hyperbolic models with nonzero growing mode (regardless of the value of the decaying mode) terminate in a line of sinks that define a future attractor {\bf MIL}, associated with the Milne spacetime contained in the {\bf VAC} invariant subspace of vacuum LTB models (see Tables 1, 2 and 4, and Figures 2 and 4). Since $\Drho\ne 0$ at {\bf MIL}, there is a terminal density inhomogeneity proportional to the amplitude of the growing mode $\Dig$, with (see section 11.4) the terminal density profile being a void ($\Dig\geq 0,\,\Drho\geq 0$) or clump ($\Dig\leq 0,\,\Drho\leq 0$). If the decaying mode is suppressed, we have the same pattern but this void/clump form of the density profile holds for the whole evolution (see section 11.4 and 12). If the growing mode is suppressed the future attractor is the Minkowski point {\bf MINK} (contained in {\bf MIL}), with $\Drho\to 0$ and thus with homogeneous terminal density (see Figure 6 and Table 4).
\item  Parabolic models are a subset of models with suppressed growing mode, evolving along the line {\bf PAR} contained in the subspace {\bf SGM}. They are described by a single trajectory that goes from the {\bf BB} past attractor of the general case towards a future attractor given by the {\bf EdS} point (see section 10, Figure 6 and Table 4). These are the only LTB models whose late time evolution approaches an Einstein de Sitter FLRW model.
\item The future attractor of trajectories of the collapsing stage of all elliptic models with nonzero decaying mode is the line of sinks {\bf BC}, associated with the Big Crunch singularity (see Figure 3, Table 2 and section 8). If the decaying mode is zero, the future attractor is the point {\bf S2} contained in {\bf BC} (see Figure 5, Table 3 and section 9). Hence, irrespective of the decaying mode, the dynamical behavior near the Big Crunch is dominated by the growing mode, and thus is qualitatively distinct from that near the Big Bang. This is an inherent feature of inhomogeneous models that is absent in FLRW re--collapsing models \cite{sussDS2}, even if the perturbations yield the same limiting values near both singularities (see section 11.3). However, the difference in behavior near these singularities becomes evident when examined in terms of the density modes and invariant scalars (see sections 11.3 and 12).    
\end{itemize}     
The invariant curvature and kinematic scalars, $\phi$ and $\zeta$, as defined by (\ref{invsc1}) and (\ref{invsc2})--(\ref{invsc3}), convey important dynamical information when we highlight their close relation with the density modes. The asymptotic early time diverging form of the ratio of Weyl to Ricci curvature $\phi$ in (\ref{ratio1}) is consistent with the notions of a ``non-isotropic'' initial singularity and ``primordial inhomogeneity'', normally associated with a dominant nonzero decaying mode, which radically changes to an ``isotropic'' initial singularity and ``primordial homogeneity'' when $\phi$ vanishes as this mode is suppressed. The late time forms (\ref{ratio2a})--(\ref{ratio2b}) are consistent with the notion of ``residual density inhomogeneity'' associated with a nonzero growing mode (see \cite{wainwright} for complementary discussion). However, notice that a nonzero decaying mode also introduces an early time primordial inhomogeneity in the Hubble scalar $\HH$ and spatial curvature $\KK$, while a nonzero growing mode does not introduce ``residual inhomogeneity'' in $\HH$ and $\KK$, since the late time vanishing of $\DKK$ and $\Dh$ implies that the invariant scalar $\zeta$ also vanishes in this limit.

The role of the density modes can also be relevant in the use of LTB models to fit cosmological observations, in particular, the preference of models with a suppressed decaying mode in void models is justified \cite{marranot} (see the dissenting view in \cite{CBK}) on the grounds that these models exhibit an early time Einstein de Sitter homogeneity (the past attractor {\bf EdS} in models with $\jd=0$, see section 9 and Figures 4 and 5). However, the demand that the decaying mode must be strictly (mathematically) zero may be too stringent, as it yields initial conditions that are too restrictive and (strictly speaking) LTB models no longer provide an appropriate description of cosmological conditions for times before the last scattering surface, and thus compatibility with observations may be achieved as long as perturbations are sufficiently small at this surface even if $\jd$ is not strictly zero (see \cite{clarkreg,zibin2,clifton}). Looking at this issue is beyond the scope of this paper and will be pursued separately.

Finally, it is important to remark that the work we have undertaken in this paper can be generalized to LTB models with nonzero cosmological constant. For this purpose the analytic solutions derived in \cite{valkenburg,romano} can be used to identify the density modes in the exact perturbation $\Drho$ (see also \cite{sussDS2}). Evidently, the case  $\Lambda\ne 0$ leads to a completely different late time dynamical behavior of the models, and this must be reflected in the exact forms of the density modes. Another necessary generalization is to non--spherical Szekeres models, proceeding along the lines of  \cite{sussbol}, and likely to even more general inhomogeneous spacetimes. These extensions of the present work are currently under elaboration and will be submitted for publication in the near future.

\begin{appendix}


\section{Review of previous literature.}

Exact expressions for the density growing and decaying mode were obtained first by Silk in \cite{silk}, which were re--derived by Krasinski and Plebanski in section 18.19 of reference \cite{kras2}, and more recently by Wainwright and Andrews in \cite{wainwright}. 

Silk (and Krasinsky--Plebanski afterwards) considered density perturbations defined by the ``comoving fractional density gradient''  $h_a^b(\rho_{,b}/\rho)=(\rho'/\rho)\delta_a^r$ \cite{ellisbruni89,BDE,1plus3} in the traditional variables (see Appendix B), which (from (\ref{fieldeqs})) involves the explicit computation of 
\begin{equation} \frac{\rho\,'}{\rho} = \frac{M'}{M}-\frac{2R'}{R}-\frac{R''}{R'},\end{equation}
for hyperbolic and elliptic models, using the parametric solutions of the Friedman equation (\ref{HHq}) to eliminate $R'$ and $R''$ in terms of $R$ and $\tbb\,M,\,E$ and their gradients. Since the resulting form of $\rho'/\rho$ is too cumbersome, these authors only examined the asymptotic limits $t\to \tbb$ (hyperbolic and elliptic models) and $t\to\infty$ (hyperbolic models) of various subexpressions in order to identify the initial conditions that suppress either one of the modes amplitudes ($\Dbe_0$ and $\tbb'$ which directly relate to $\Dig$ and $\Did$). Notice that the parameters $\alpha,\,\beta,\,\gamma$ of \cite{kras2} respectively correspond to $(3/r)(\Dal_0-1/3),\,(-3/r)\Dbe_0,\,-1/\beta_{q0}$ defined in (\ref{Dbe0}) and (\ref{Dal0}). There is no attempt in \cite{silk} or \cite{kras2} to obtain the full expressions of the modes or to study their relation with generic properties of the models (regularity conditions or density profiles). As a contrast, we have used the density perturbation $\Drho$ in (\ref{perts1}) and (\ref{DrhoJ}), related to the gradient $\rho'_q/\rho_q$ by (\ref{Dadef}), which allows us to identify the same amplitudes and yields much more tractable subexpressions for the modes (just compare the elegance and simplicity of (\ref{gmode}) and (\ref{dmode}) with the rather awkward equations of section 18.19 of \cite{kras2}).

Wainwright and Andrews considered exactly the same metric (\ref{ltb2})--(\ref{aGdef}) [their equation (2.1)], with $a$ satisfying the Friedman equation (\ref{HHq}) [their equation (2.5)] with $\Lambda>0$ and their parameters $m,\,k$ respectively corresponding to  $(8\pi/3)\rho_{q0},\,\KK_{q0}$. They used the ansatz $\Gamma=1+\Delta$, with the  ``deviation function'' $\Delta$ given in the Goode--Wainwright form $\Delta=\beta_+( r)f_+(t,r)+\beta_-( r)f_-(t,r)$, which (according to the authors) identifies the growing ($+$) and decaying ($-$) modes. 

However, Wainwright and Andrews assumed without any justification that their parameter $m$ ({\it i.e.} $\rho_{q0}$) is constant, which (in general) is not true, and thus their study removes unjustifiably an important degree of freedom in the set of initial conditions. As a consequence, the following key results are only valid for the rather uninteresting particular case of models admitting a constant initial density ($\Drho_0=0$):
\begin{itemize}
\item The relation between $\Delta^2$ and the ratio of quadratic invariant scalar contractions of the Weyl and Ricci tensors [their equation (2.8)] is not valid in general. If we consider the full degrees of freedom $\Delta^2$ does not comply with such ratio, taking instead the form:
\begin{equation} \Delta^2 =\left[\frac{6\Psi_2-\Drho_0(\RRR-6\Psi_2)}{\RRR}\right]^2,\end{equation}
where $\Psi_2$ is the only nonzero conformal invariant in a Newman--Penrose representation and $\RRR$ is the 4--dimensional Ricci scalar. Evidently, if $\Drho_0=0$ we recover the equation (2.8) of Wainwright and Andrews, as the scalar contractions $C_{abcd}C^{abcd}$ and $\RRR_{ab}\RRR^{ab}$ are respectively proportional to $\Psi_2^2$ and $\RRR^2$ (see \cite{part1}).
\item The evolution equation for $\Delta$ [their equation (2.9)] is in general given by:
\begin{equation}a\dot a\dot\Delta +\left(\frac{3m}{a^3}-k\right)\Delta = \frac{r}{2}\left[k'( r) -\frac{2m'( r)}{a}\right],\end{equation}
and reduces to (2.9) of Wainwright and Andrews only if $m'=0$ (equivalent to $\Drho_0=0$). As a consequence, the general evolution equation for $\ddot \Delta$ is not given by (2.10).
\item Equation (3.5) is not general. Hence, following these authors, if we define the growing/decaying modes from $\Delta$ in the general case we obtain
\begin{equation} \fl \Delta = 1-\Gamma = r\tbb'\frac{\dot a}{a}-\frac{r}{2}\frac{\dot a}{a}\,I,\qquad I =\int_0^a{\frac{\left[2m'-k'\,\bar a\right]\bar a^{1/2}}{\left[2m-k\,\bar a+\frac{1}{3}\Lambda\,\bar a^3\right]^{3/2}}}, \end{equation}
which yields their equations (3.7) and (3.8) only if $m'=0$. However, while the forms for $\beta_-$ and $f_-$ in (3.7) and (3.8) do coincide with (\ref{DT0}) and $\HH_q=\dot a/a$, and thus allow us to express the decaying mode (\ref{dmode}) as the product $\beta_-( r)f_-(t,r)$, it is not possible (in general) to express the term $(\dot a/a)\,I$ as a product $\beta_+( r)f_+(t,r)$ that would allow for the identification of the growing mode (\ref{gmode}) and its amplitude (\ref{Dbe0}).
\end{itemize}
Since the assumption ``$m'=0$'' carries along the rest of their paper, some (or a lot) of the results of Wainwright and Andrews may be misleading or even mistaken. In particular, a decomposition of the growing/decaying modes in terms of the Goode--Wainwright variables $\beta_\pm\,f_\pm$ obtained form the metric function $\Delta$ does not seem to be possible in general. In contrast, such decomposition leading to the correct linear limit is possible and consistent through the density perturbation $\Drho$ in  (\ref{DrhoJ}). Besides the issue of consistency, we remark that $\Drho$ is not a metric function, but a coordinate independent quantity related by (\ref{invsc2}) to the ratio of invariant scalars $\Psi_2$ and $\RRR$.             

\section{LTB models in their standard variables.}

LTB models are usually described in their original variables by the metric and field equations:
\ba \dd s^2 = -\dd t^2 + \frac{R'{}^2}{1+2E}\,\dd r^2+R^2\left(\dd\vartheta^2+\sin^2\vartheta\,\dd\varphi^2\right),\label{ltbold}\\
\dot R^2 = \frac{2M}{R}+2E,\qquad 4\pi\rho =\frac{M'}{R^2R'}\label{fieldeqs}
\ea
with the analytic solutions of the Friedman--like equation above usually given in parametric form (though these solutions can also be given in the implicit forms (\ref{hypsol})--(\ref{ellsol})).  

We can obtain the metric (\ref{ltb2}) and the q--scalar variables from (\ref{ltbold}) and (\ref{fieldeqs}) by selecting the radial coordinate so that $R_0=R(t_0,r)=r$ and re--scalling $R,\,M,\,E$ as 
\ba \fl \frac{R}{r}= a,\quad 
\frac{2M}{r^3}=\frac{8\pi}{3}\rho_{q0}=\Omega_{q0}\HH_{q0}^2,\quad \frac{2E}{r^2}=-\KK_{q0}=(1-\Omega_{q0})\HH_{q0}^2,\label{rescalings1}\\
\fl \Omega_{q0}=\frac{M}{M+Er},\qquad \HH_{q0}=\frac{[2\,(M+Er)]^{1/2}}{r^{3/2}},\label{rescalings2}\ea
with $\Gamma=rR'/R$ and $\HH_q=\dot R/R=\dot a/a$ given in terms of $R,\,M,\,E,\,\tbb$ by 
\ba\fl \frac{\Gamma}{r} = \frac{M'}{M}-\frac{E'}{E}-\frac{3\dot R}{R}\left[(t-\tbb)\left(\frac{M'}{M}-\frac{3E'}{2E}\right)+\tbb'\right],\quad 
\frac{\dot R}{R} =\frac{[2(M+ER)]^{1/2}}{R^{3/2}}.\label{oldnot}\ea
All the results of this article can be immediately re-written in terms of the variables $R,\,M,\,E,\,\tbb$ by direct substitution of (\ref{rescalings1}), (\ref{rescalings2}) and (\ref{oldnot}) in the appropriate scaling laws and analytic expressions.  
  
\end{appendix}

\end{document}